\documentclass[12pt]{JHEP3}

\def\be{\begin{equation}}
\def\ee{\end{equation}}

\title{Abelian Toda field theories on the noncommutative plane}
\author{I. Cabrera-Carnero\\Departamento de F\'{\i}sica-ICET \\
Universidade Federal de Mato Grosso \\
Av. Fernando Corr\^{e}a da Costa, $s/n^o$-Bairro Coxip\'{o}\\
78060-900-Cuiab\'{a}-MT-Brasil \\ \email{Email:
cabrera@cpd.ufmt.br}}

\abstract{Generalizations of $GL(n)$ abelian Toda and
$\widetilde{GL}(n)$ abelian affine Toda  field theories to the
noncommutative plane are constructed. Our proposal relies on the
noncommutative extension of a zero-curvature condition satisfied
by algebra-valued gauge potentials dependent on the fields. This
condition can be expressed as noncommutative Leznov-Saveliev
equations which make possible to define the noncommutative
generalizations as systems of second order differential equations,
with an infinite chain of conserved currents. The actions
corresponding to these field theories are also provided. The
special cases of $GL(2)$ Liouville and $\widetilde{GL}(2)$
sinh/sine-Gordon are explicitly studied. It is also shown that
from the noncommutative (anti-)self-dual Yang-Mills equations in
four dimensions it is possible to obtain by dimensional reduction
the equations of motion of the two-dimensional models constructed.
This fact supports the validity of the noncommutative version of
the Ward conjecture. The relation of our proposal to previous
versions of some specific Toda field theories reported in the
literature is presented as well. }

\begin{document}

\section{Introduction}

The research on Noncommutative Field theories (NCFT) has been very
active since the appearance of these theories as low-energy limits
of string theories in the presence of magnetic fields
\cite{stringnc}. In the context of NCFT, noncommutative (nc)
extensions of two-dimensional Integrable Field Theories have been
investigated \cite{dimakis, integrablenc, integrablenc2, GP1, us,
GP2, lechtenfeld, kimeon, zuevsky}. Since in two-dimensions a nc
deformation of a model requires a noncommutative time-coordinate,
the causality and unitarity properties of the theory can be
compromised \cite{GM, BDFP}. However, it is conceivable that in
exactly solvable systems this situation should be improved or even
disappear, as discussed in \cite{CLZ}. In order to avoid the
acasual behavior in the two-dimensional case, Euclidean models can
be considered.

It is well known that the nc deformation of a theory is not unique
since it is always possible to construct different nc extensions
that will lead to the same commutative limit (see the nc
generalizations of sine-Gordon model in \cite{us}). In this sense,
preserving the integrability properties of the theory can be a
guiding principle in order to construct nc deformations of
two-dimensional theories.

Following the previous direction, in \cite{us} a nc extension of
the zero-curvature condition was introduced. The nc extensions of
integrable theories, constructed from this condition, have an
infinite number of conserved charges which, however, not always
guaranteed the complete classical integrability of the theory (see
\cite{GP2}). The amplitude for particle production processes
vanishes exactly in an integrable model and that means that it
vanishes to each order in a loop expansion, that is, in powers of
$\hbar$. In particular it should vanish at tree-level, which
corresponds to the classical limit of the theory, and is the
hallmark of classical integrability. The existence of non-trivial
conserved charges has another important consequence:
multi-particle amplitudes are factorized into products of two-body
processes \cite{D}. In \cite{GP2} was proved that the nc extension
of the sine-Gordon model constructed in \cite{GP1,us} from a nc
zero-curvature condition suffers from acasual behavior and it has
a non-factorized S-matrix since particle production occurs.
Therefore, we can be sure that this model is not classically
integrable.

In the ordinary commutative case, it is well known that the
integrable Conformal Toda (and Conformal Affine Toda) field
theories can be obtained from the Wess-Zumino-Novikov-Witten
($WZNW$) \cite{witten} (two-loop $WZNW$) mo\-del via Hamiltonian
Reduction \cite{balog, aratyn}. The algebraic structure of Toda
theories is connected with a $G_0 \in G$ embedding of the
$G$-invariant $WZNW$ (two-loop $WZNW$) model. Specifically abelian
Toda theories are connected with abelian $G_0$ subgroups of $G$.
In order to eliminate the degrees of freedom in the tangent space
$G/G_0$ it is possible to implement constraints upon specific
components of the chiral currents $J, \bar{J}$ of the $WZNW$
(two-loop $WZNW$) model. The equations of motion of the resultant
model will be then defined in the $G_0$ subgroup. Usually these
equations can be represented as a zero-curvature condition using
the Leznov-Saveliev formulation \cite{LezSav}. The resultant Toda
theory preserves also the original conformal symmetry of the
$WZNW$ model. By the other side, the affine Toda models associated
to loop algebras are not conformally invariant but it has been
shown to be completely integrable \cite{olive} and also derivable
from the Leznov-Saveliev equations \cite{LezSav}. These models are
in fact a ``gauge fixed" version of the Conformal Affine Toda
models \cite{constantinidis}.

In this paper we use a nc zero-curvature condition expressed as a
nc extension of Leznov-Saveliev equations \cite{us} to construct
nc integrable extensions of $SL(n)$ abelian Toda field theories
and $\widetilde{SL}(n)$ abelian affine Toda field theories. In
order to define the zero grade subgroup $G_0$ we have taken into
account that the previous groups are not closed under the
noncommutative product, so they should be extended to $GL(n)$ and
$\widetilde{GL}(n)$ respectively. If we want to preserve the
proper {\it algebra-group} relation this extension must be also
done at the level of the algebra. This consideration imply the
introduction of an additional scalar field associated to the
identity generator and which will not decouple in the equations of
motion. We explicitly studied the $GL(2)$ Liouville and
$\widetilde{GL}(2)$ sinh/sine-Gordon extensions. As we will see,
any element of the zero grade subgroup can be parameterized in
alternative ways, what leads to equivalent nc generalizations of
the corresponding original models. This is a consequence of the
nonabelian cha\-rac\-ter of the zero grade subgroup in the nc
setup. The models obtained as nc analogs of sine-Gordon reproduce
already presented suggestions in \cite{lechtenfeld} as nc
extensions of this model. These generalizations \cite{lechtenfeld}
seem to retain some of the nice properties of the original
sine-Gordon theory. They have an infinite chain of conserved
charges, apparently a casual S-matrix, and particle production may
be does not occur as was checked for some dispersion processes at
tree level \cite{lechtenfeld}.

Another argument that supports the possible integrability of the
nc Toda field theories constructed in this work is that all of
them can be derived from the $4$-dimensional nc self-dual
Yang-Mills (SDYM) equations through a suitable dimensional
reduction via the nc self-dual Chern-Simons (SDCS) system as we
will see. The nc SDYM theory was shown in \cite{takasaki} to be
classically integrable and in \cite{popov3} was studied at the
quantum level, where it was found that it has a factorized
S-matrix. In this work we will also particularly see how the
equations of motion that define the nc extensions of Liouville and
sinh/sine-Gordon models can be obtained through a dimensional
reduction process from $4$-dimensional nc self-dual Yang-Mills in
the Yang formulation. In this way all the models are derivable
from an integrable field theory.

On the other hand, the nc extensions here constructed do not
preserve the conformal invariance of the original abelian
Conformal Toda theories. As it is well known, the introduction of
a constant noncommutative parameter spoils the conformal symmetry.
This symmetry has not been very well studied in the nc context,
but it seems that in order to define a nc extension of this
symmetry the deformation parameter should not be constant
\cite{pinzul}.

In the literature it has been reported different nc extensions of
some specific Toda field theories: 1-sinh/sine-Gordon
\cite{GP1,GP2,us, lechtenfeld, kimeon,zuevsky}, 2-Liouville
\cite{kimeon,zuevsky}, 3-open Toda field chain or abelian
Conformal Toda field theories \cite{dimakis,kimeon}, 4- abelian
affine Toda field theories or closed Toda field chain
\cite{kimeon}. Up to now many of these extensions for the same
models seem to be disconnected. With our formulation we are given
a general framework where all these models are included as
particular cases, excluding the proposals in \cite{zuevsky}.
Therefore we have extended these previous results, putting most of
them on a more systematic, general and unifying footing.

 The first nc extension of a Toda
field theory was presented in \cite{dimakis}. It is well known
that the abelian Conformal Toda theories when the fields only
depend on the time coordinate can be modelled as a one-dimensional
open chain of $n$ particles with nonlinear nearest neighbor
interaction \cite{Toda}. The abelian affine corresponds to closed
chains. Deforming the bicomplex representation of the open Toda
field chain it was constructed in \cite{dimakis} a nc extension of
this theory up to first order in the noncommutative parameter
$\theta$. Here we will see how our nc extension of abelian Toda
models reduces to the proposal in \cite{dimakis} when a
perturvative expansion on the noncommutative parameter is
considered up to first order. By the other side in \cite{kimeon}
it was shown that using some simplifying algebraic ansatz the
two-dimensional nc self-dual equations for the Chern-Simons
solitons can be reduced to a nc extension of the Toda and affine
Toda equations. The nc generalization for the Toda field theories
proposed in \cite{kimeon} is presented as a system of first order
differential equations that apparently could not be reduced to
second order differential equations. In this work we will see how
proposing a different ansatz it is possible to construct from the
nc Chern-Simons self-dual system extended Toda field theories as
second order differential equation systems and even to establish
the compatibility with the suggestion in \cite{kimeon}.
 In this sense with our proposal we have
nc versions of abelian and abelian affine Toda theories more
treatable as physical theories since we also provide the
corresponding actions. Moreover it is still possible to use the
relation of the nc self-dual Chern-Simons equations to the nc
chiral model for constructing the solutions of the models.
Specifically this last point will be presented in \cite{sdchsyo}.
There are other nc versions for the specific cases of Liuoville
and sine-Gordon \cite{zuevsky}. In this case the construction is
based on a zero-curvature condition in terms of continual algebras
\cite{save-ver}. A brief discussion about these models is also
presented in this paper although they can not be obtained using
our formalism. Let us mention that in \cite{sakakibara} was also
extended to the nc scenario the Toda hierarchy which is a
generalization of the Toda lattice equations \cite{ueno}.

This paper is organized as follows. In the first section the nc
Leznov-Saveliev equations are obtained by imposing appropriate
constraints on the chiral currents of the $WZNW_{\star}$ model. In
this section we also provide a nc action whose Euler-Lagrange
equations of motion are the nc Leznov-Saveliev equations. Section
$2$ is devoted to the construction of the nc analogs of $GL(n)$
abelian Toda theories. Specially the relation to the nc open Toda
field theory presented in \cite{dimakis} is discussed. The
particular case of $GL(2)$ Liouville model is stu\-died in the
third part of this section. The fourth part contains the
derivation of the equations of motion of nc Liouville from nc self
dual Yang-Mills in the Yang formulation \cite{takasaki}. In the
last part of section $3$ the relation to previous proposals is
discussed. Starting from the nc extension of the Leznov-Saveliev
equations we construct in section $4$ the nc generalization of
$\widetilde{GL}(n)$ abelian affine Toda theories. In this section
is also studied the special case of $\widetilde{GL}(2)$
sinh/sine-Gordon model. The derivation of its corresponding
equations of motion from nc self dual Yang-Mills is presented as
well. At the end of section $4$ the relation to previous versions
is included. In the last section it is shown how the nc
Leznov-Saveliev equations \cite{us} can be derived from the nc
(anti-)self-dual Yang-Mills theory by a dimensional reduction
process that has the nc self-dual Chern-Simons system as an
intermediate step. In this sense the connection of our nc
extensions of abelian and abelian affine Toda theories to the
proposal in \cite{kimeon} is established. Section $6$ provides the
conclusions and finally the appendix is dedicated to present some
useful algebraic properties.

\section{Constrained $WZNW_{\star}$ model}

It is well known that Toda field theories connected with finite
simple Lie algebras, on the ordinary commutative case, can be
regarded as constrained Wess-Zumino-Novikov-Witten ($WZNW$) models
\cite{balog}. By placing certain constraints on the chiral
currents, the G-invariant $WZNW$ model reduces to the appropriate
Toda field theory. Specifically, the abelian Toda field theories
are connected with abelian embeddings $G_0\subset G$. In this
section we will see how this procedure also works on the nc
setting.

Before we start, let us remind that usually a NCFT \cite{reviews}
is constructed from a given field theory by replacing the product
of fields by an associative $\star$-product. Considering that the
noncommutative parameter $\theta^{\mu \nu}$ is a constant
antisym\-me\-tric tensor, the deformed product of functions is
expressed trough the Moyal product \cite{moyal} \be
  \phi_1(x)\phi_2(x) \to
\phi_1(x)\star\phi_2(x)=e^{\frac{i}{2}\theta^{\mu\nu}\partial_\mu^{x_1}
\partial_\nu^{x_2}}\phi_1(x_1)\phi_2(x_2)|_{x_1=x_2=x}.
\ee In the following we will refer to functions of operators in
the noncommutative deformation by a $\star$ sub-index, for example
$e^{\phi}_\star=\sum_{n=1}^{\infty}\frac{1}{n!}\phi_\star^{n}$
(the n-times star-product of $\phi$ is understood).

 Consider now the nc generalization of the $WZNW$ model introduced
in \cite{wznwnc}
\begin{eqnarray}
S_{WZNW_{\star}}&=&-\frac{k}{4\pi}\int_{\Sigma} d^2z Tr (
g^{-1}\star
\partial{g} \star g^{-1} \star \bar{\partial}g)+ \nonumber \\
&&\frac{k}{24\pi}\int_{{\cal B}} d^3 x \epsilon_{ijk}(g^{-1}\star
\partial_i g \star g^{-1}\star
\partial_j g \star g^{-1}\star \partial_k g ). \label{wznwnc}
\end{eqnarray}
Here ${\cal B}$ is a three-dimensional manifold whose boundary
$\partial {\cal B}=\Sigma $. We are using the coordinates
 $z=t+x$, $\bar{z}=t-x$ and
$\partial=\frac{1}{2}(\frac{\partial}{\partial
t}+\frac{\partial}{\partial
 x})$, $\bar{\partial}=\frac{1}{2}(\frac{\partial}{\partial
t}-\frac{\partial}{\partial
 x})$ in the boundary, where $z, \bar{z}$ or
equivalently $x,t$ are noncommutative, but the extended coordinate
$y$ on the manifold ${\cal B}$ remains commutative, i.e.
$[z,\bar{z}]=\theta$, $[y,z]=[y,\bar{z}]=0$. The Euler-Lagrange
equations of motion corresponding to (\ref{wznwnc}) are
\begin{eqnarray}
 \bar{\partial}J=\partial \bar{J}=0, \label{eqnwznwnc}
 \end{eqnarray}
where $J$ and $\bar{J}$ represent the conserved chiral currents
\begin{eqnarray}
J=g^{-1}\star \partial g,  \quad \quad \quad
\bar{J}=-\bar{\partial}g \star g^{-1}. \label{chiralcurrents}
\end{eqnarray}
The fields $\alpha_a$ parameterize the group element $g \in G$
through $ g=e_{\star}^{\alpha_a T_a}$,
  where $T_a$ are the generators of the corresponding Lie algebra ${\cal
G}$. It is our interest to define the theories in a $G_0$ subgroup
of $G$. So, we would like to eliminate the unwanted degrees of
freedom that correspond to the tangent space $G/G_0$. In order to
that be achieved we will implement constraints upon specific
components of the currents $J, \bar{J}$. In this way we will show
that the usual procedure \cite{balog} works equally well on the nc
setting.

Suppose we have defined a grading operator $Q$ in the algebra
$\cal{G}$ that decomposes it in $\mathbb{Z}$-graded subspaces, say
 \begin{eqnarray}
[Q,{\cal G}_i]=i{\cal G}_i, \quad \quad  [{\cal G}_i,{\cal G}_j]
\in {\cal G}_{i+j}.
 \end{eqnarray}
This means that the algebra ${\cal G}$ can be represented as the
direct sum,
 \be
 {\cal G}=\bigoplus_i {\cal G}_i.
 \ee
The subspaces ${\cal G}_0,{\cal G}_>,{\cal G}_{<}$ are subalgebras
of ${\cal G}$, composed of the Cartan and of the positive/negative
steps generators respectively.
 The algebra can then
be written using the triangular decomposition,
 \be
 {\cal G}={\cal G}_<\bigoplus {\cal G}_0 \bigoplus {\cal G}_> .
\ee Denote the subgroup elements obtained through the
$\star$-exponentiation of the ge\-ne\-ra\-tors of the
corresponding subalgebras as
\begin{eqnarray}
N=e^{{\cal G}_<}_{\star}, \quad B=e^{{\cal G}_0}_{\star}, \quad
M=e^{{\cal G}_>}_{\star}.
\end{eqnarray}
Proposing a nc Gauss-like decomposition, an element $g$ of the nc
group $G$ can be expressed as
 \be
 g=N \star B \star M. \label{Gaussnc}
 \ee
Introducing (\ref{Gaussnc}) in (\ref{chiralcurrents}), the chiral
currents $J, \bar{J}$ read,
\begin{eqnarray}
J=M^{-1}\star K \star M, \quad \quad \bar{J}=-N\star \bar{K}\star
N^{-1}, \label{JJ}
\end{eqnarray}
where
\begin{eqnarray}
K&=&B^{-1} \star N^{-1} \star \partial N \star B+B^{-1} \star
\partial B
+\partial M \star M^{-1},\nonumber \\
\bar{K}&=&N^{-1}\star \bar{\partial} N +\bar{\partial} B \star
B^{-1}+ B \star \bar{\partial} M \star M^{-1}\star B^{-1}.
\label{KK}
\end{eqnarray}
With the chiral currents (\ref{JJ}), the equations of motion
(\ref{eqnwznwnc}) transform to
\begin{eqnarray}
\bar{\partial}K+[K,\bar{\partial}M\star M^{-1}]_{\star}&=&0,
\nonumber
\\
\partial \bar{K}-[\bar{K}, N^{-1} \star \bar{\partial}N]_{\star}&=&0.
\label{kkeq}
\end{eqnarray}
 The reduced model is defined by giving the constant elements
$\epsilon_{\pm}$ of grade $\pm1$, which are responsible for
  cons\-training the currents in a general manner to \footnote{See
\cite{balog} for the commutative case.}
\begin{eqnarray}
J_{constr}=j+\epsilon_-, \quad \quad \quad
\bar{J}_{constr}=\bar{j}+\epsilon_{+}, \label{HRconstraints}
\end{eqnarray}
where $j,\bar{j}$ contain generators of grade zero and positive,
and zero and negative res\-pec\-ti\-vel\-y. The effect of the
constraints on the chiral currents $J, \bar{J}$ (\ref{JJ})
translates in the conditions,
\begin{eqnarray}
B^{-1} \star N^{-1}\star \partial N \star
B|_{constr}&=&\epsilon_-,
\nonumber \\
B \star \bar{\partial} M \star M^{-1}\star
B^{-1}|_{constr}&=&\epsilon_+,
\end{eqnarray}
because from the graded structure these are the only terms in
(\ref{KK}) that contain gene\-ra\-tors of negative and positive
grade respectively. As result of the reduction process, the
degrees of freedom in $M,N$ are eliminated and the equations of
motion of the constrained model are natural nc extensions of the
Leznov-Saveliev equations of motion \cite{LezSav}, namely
\begin{eqnarray}
\bar{\partial}(B^{-1}\star\partial
B)+[\epsilon_-,B^{-1}\star\epsilon_+ B]_{\star}=0,
\nonumber \\
\partial(\bar{\partial}B \star B^{-1})-[\epsilon_+,B \epsilon_-\star
B^{-1}]_{\star}=0. \label{Lez-Savnc}
\end{eqnarray}
Both equations are equivalent. One can see that by
$\star$-multiplying $B$ from the left and $B^{-1}$ from the right
the first equation in (\ref{Lez-Savnc}), say
\begin{eqnarray}
B\star\{\bar{\partial}(B^{-1}\star\partial
B)+[\epsilon_-,B^{-1}\star\epsilon_+ B]_{\star}&=&0\}\star B^{-1},
\nonumber \\ -\bar{\partial} B \star B^{-1} \star \partial B \star
B^{-1}+\bar{\partial} \partial B\star B^{-1}-[\epsilon_+,B
\epsilon_-\star B^{-1}]_{\star}&=&0.
\end{eqnarray}
Then, using that $\partial(B \star B^{-1})=0$ the second equation
in (\ref{Lez-Savnc}) is obtained, what means that both equations
are simultaneously satisfied. In \cite{us} these equations were
used to define a nc extension of the sinh/sine-Gordon
 model. In contrast to the previous suggestion \cite{us}, in this paper
we propose an
 alternative definition
 for $B$ which preserves the proper {\it algebra-group} relation.
 This choice will lead to a nc sine-Gordon defined as a system of
two coupled
 second order equations for two scalar fields that reduced to
sine-Gordon
 model and a free scalar field in the commutative limit.
 The nc extensions of the Leznov-Saveliev equations
 (\ref{Lez-Savnc}) for $GL(2)$
 were also
 obtained in \cite{harold} from the nc generalization of the $SL(2)$
affine Toda model coupled to matter (Dirac)
 fields.

As shown in \cite{us}, the equations of motion (\ref{Lez-Savnc})
can be expressed as a generalized $\star$-zero-curvature condition
 \be
 \bar{\partial}A-\partial
\bar{A}+[A,\bar{A}]_{\star}=0, \label{starzc} \ee since the
potentials are taken as
\begin{eqnarray}
A=-B \epsilon_-\star B^{-1}, \quad \quad
\bar{A}=\epsilon_++\bar{\partial}B \star B^{-1}.
\label{potentialnc}
\end{eqnarray}
The $\star$-zero-curvature condition (\ref{starzc}) implies the
existence of an infinite amount of conserved charges \cite{us}.
For this reason in order to preserve the original integrability
properties of the two-dimensional Toda models (\ref{Lez-Savnc})
can be a reasonable starting point for constructing the nc
analogs.

\subsection{The noncommutative action}

It is not difficult to propose a nc action from which the nc
Leznov-Saveliev equations (\ref{Lez-Savnc}) can be derived. In
fact, (\ref{Lez-Savnc}) are the Euler-Lagrange equations of motion
of the action
 \be
S=S_{WZNW_{\star}}(B)+\frac{k}{2\pi}\int d^2z Tr (\epsilon_+ B
\epsilon_-\star B^{-1}). \label{effectiveactionnc}
 \ee
This is the nc generalization of the effective action obtained
from the $WZNW$ model gauging the degrees of freedom in $M,N$ and
integrating over the corresponding gauge fields \cite{aratyn}. In
\cite{parvizi} different gauged $WZNW_{\star}$ models were
constructed. Howe\-ver, the integration over the gauge fields on
the nc scenario requires special care. For this reason in the
present paper we limit to propose the action
(\ref{effectiveactionnc}) as corresponding to the equations
(\ref{Lez-Savnc}) and it remains to be proved if
(\ref{effectiveactionnc}) can be obtained from (\ref{wznwnc}) by a
gauging procedure.

\section{NC $GL(n)$ abelian Toda field theories}

In the following we will construct nc analogs of the $SL(n)$
abelian Toda field theories, the simplest class of Toda models
which are completely integrable and conformal invariant. As we
already mentioned, these models correspond to an abelian subgroup
$G_0 \subset G$. Considering the extension of the ${\cal SL}(n)$
algebra to ${\cal GL}(n)$, what is necessary in order to obtain a
nc closed group, the gradation operator \be
 Q=\sum_{i=1}^{n-1}\frac{2\lambda_i
\cdot H}{\alpha_i^2}, \label{grading2}
 \ee
defines the subalgebra of grade zero ${\cal
G}_0=U(1)^{n}=\{I,h_i,\, \,\,i=1 \dots n-1\}$, where the Cartan
generators are defined in the Chevalley basis as $h_i=\frac{2
\alpha_i \cdot H}{\alpha_i^2}$. In (\ref{grading2}) $H$ represents
the Cartan subalgebra, $\alpha_i$ is the $i^{th}$ simple root and
$\lambda_i$ is the $i^{th}$ fundamental weight that satisfies $
\frac{2\lambda_i \cdot \alpha_j}{\alpha_i^2}=\delta_{ij}$. The
zero grade group element $B$ is then expressed through the
$\star$-exponentiation of the generators of the zero grade
subalgebra ${\cal G}_0$, i.e. the ${\cal SL}(n)$ Cartan subalgebra
plus the identity generator,
\begin{eqnarray}
B=e^{\Sigma_{i=1}^{n-1}\varphi_i h_i + \varphi_0 I}_{\star}.
\label{B}
\end{eqnarray}
Notice that the zero grade subgroup $G_0$ despite it is spanned by
the generators of the Cartan subalgebra, turns out to be
nonabelian, i.e, if $g_1,g_2$ are two elements of the zero grade
subgroup $G_0$ then $g_1 \star g_2 \neq g_2 \star g_1$. For this
reason {\it abelian} makes reference to the property of the
original theory.

The constant generators of grade $\pm1$ are chosen as
\begin{eqnarray}
\epsilon_{\pm}= \Sigma_{i=1}^{n-1}\mu_i E_{\pm \alpha_i},
\label{eps}
\end{eqnarray}
where $E_{\pm \alpha_i}$ are the steps generators associated to
the positive/negative simple roots of the algebra and $\mu_i$ are
constant parameters.

Let us consider the $n \times n$ matrix representation
\begin{eqnarray}
(h_i)_{\mu\nu}=\delta_{\mu\nu}(\delta_{i, \mu}-\delta_{i+1, \mu}),
& (E_{\alpha_i})_{\mu \nu}= \delta_{\mu, i}\delta_{\nu, i+1}, &
(E_{-\alpha_i})_{\mu \nu}= \delta_{\nu, i}\delta_{\mu, i+1}.
\label{representation}
\end{eqnarray}
It is not difficult to see that in this case the zero grade group
element (\ref{B}) can be represented by the $n \times n$ diagonal
matrix,
 \be
B=\left(\begin{array}{cccccc}
e_{\star}^{\varphi_1+\varphi_0}& 0& 0& 0& \dots & 0\\
0& e_{\star}^{-\varphi_1+\varphi_2+\varphi_0}& 0& 0& \dots & 0\\
0& 0& e_{\star}^{-\varphi_2+\varphi_3+\varphi_0}& 0& \dots & 0\\
0&0&0 & \ddots & \ddots & \vdots \\
\vdots & \vdots & \vdots & \ddots &
e_{\star}^{\varphi_{n-1}-\varphi_{n-2}+\varphi_0} & 0 \\
0& 0 & 0& \dots & 0& e_{\star}^{-\varphi_{n-1}+\varphi_0}
        \end{array}\right). \label{Bnctoda}
\ee Let us now introduce the variables,
\begin{eqnarray}
\varphi_1+\varphi_0&=&\phi_1, \nonumber \\
-\varphi_k+\varphi_{k+1}+\varphi_0&=&\phi_{k+1}, \, \, \, \,
\mathrm{for} \, \, k=1 \, \, \mathrm{to} \, \, n-2,
\label{newvariables}
\\
-\varphi_{n-1}+\varphi_0&=&\phi_n. \nonumber
\end{eqnarray}
In these new fields the components of the gauge connections
(\ref{potentialnc}) are written as
\begin{eqnarray}
\bar{A}_{ij}=\bar{\partial}(e_{\star}^{\phi_i})\star
e_{\star}^{-\phi_i}\delta_{ij}+\mu_i \delta_{i+1,j} \quad \mathrm{
and} \quad A_{ij}=-\mu_ie_{\star}^{\phi_{i+1}}\star
e^{-\phi_i}_{\star}\delta_{i,j+1}, \label{todapotentials}
\end{eqnarray}
with $i,j=1\dots n$. One can now introduce the gauge potentials
(\ref{todapotentials}) in the $\star$-zero-curvature condition
(\ref{starzc}) to obtain the equations of motion that define the
nc extension of abelian Toda models,
\begin{eqnarray}
\partial(\bar{\partial}( e_{\star}^{\phi_{k}})\star
e_{\star}^{-\phi_{k}})&=& \mu_{k}^2 e_{\star}^{\phi_{k+1}}\star
e_{\star}^{-\phi_{k}}-\mu_{k-1}^2e_{\star}^{\phi_{k}}\star
e_{\star}^{-\phi_{k-1}},
 \label{todaeqnmotion}
\end{eqnarray}
a system of $n$-coupled equations ($k=1\dots n$). Notice that the
diagonal elements of the matrix equation (\ref{starzc}) are the
equations of motion and the off-diagonal elements vanish. In
(\ref{todaeqnmotion}) for the first and last equation
$\mu_0=\mu_n=0$ and $\phi_0=\phi_{n+1}=0$. In order to compute the
commutative limit we introduce the original fields $\varphi_0,
\varphi_1,\dots \varphi_{n-1}$ (\ref{newvariables}) and then we
apply the limit $\theta \rightarrow 0$, which transforms
$e_{\star}^{\phi}\rightarrow e^{\phi}$. At the end we find, as
expected, that the field $\varphi_0$ decouples, i.e.
(\ref{todaeqnmotion}) leads to
 \begin{eqnarray}
\partial \bar{\partial}
\varphi_i&=&\mu_i^2 e^{-k_{ij}\varphi_j}, \quad \quad \mathrm{for}
\quad i=1\dots n-1, \nonumber \\
n \partial \bar{\partial} \varphi_0 &=&0, \label{abeliantodaeq}
\end{eqnarray}
where the first $n-1$ equations become the usual abelian Conformal
Toda equations with $k_{ij}$ the Cartan matrix.

The action, whose Euler-Lagrange equations of motion leads to
(\ref{todaeqnmotion}), can be obtained from
(\ref{effectiveactionnc}) with (\ref{eps}) and (\ref{Bnctoda}). It
reads
\begin{eqnarray}
S(\phi_1,\dots,\phi_n)&=&
\sum_{k=1}^{n}S_{WZNW_{\star}}(e_{\star}^{\phi_k})+\frac{k}{2\pi}\int
d^2z \sum_{k=1}^{n-1}\mu^2_k(e_{\star}^{\phi_{k+1}}\star
e_{\star}^{-\phi_{k}}), \label{todancaction2}
\end{eqnarray}
that in the commutative limit yields
\begin{eqnarray}
S(\varphi_1,\dots,\varphi_{n-1},\varphi_0)&=&S_{CT}(\varphi_1,\dots,\varphi_{n-1})+nS_0(\varphi_0)
\end{eqnarray}
where
\begin{eqnarray}
 S_{CT}(\varphi_1,\dots,\varphi_{n-1})&=&-\frac{k}{4\pi}\int d^2z
(k_{ij}\partial \varphi_i \bar{\partial}
\varphi_j-2\sum_{i=1}^{n-1}\mu_i^2
e^{-k_{ij}\varphi_j}), \nonumber \\
S_0(\varphi_0)&=&-\frac{k}{4\pi}\int d^2z \partial \varphi_0
\bar{\partial} \varphi_0. \label{abeliantoda}
\end{eqnarray}
This is the action of the abelian Conformal Toda models plus the
corresponding kinetic term for the free field $\varphi_0$. In the
last calculation we have made use of the nc generalization of the
Polyakov-Wiegmann identity
\begin{eqnarray}
S_{WZNW_{\star}}(g_1\star
g_2)&=&S_{WZNW_{\star}}(g_1)+S_{WZNW_{\star}}(g_2)-\nonumber \\
& & -\frac{1}{2\pi}\int dzd\bar{z}Tr(g_1^{-1}\star \partial g_1
\star \bar{\partial} g_2 \star g_2^{-1}).
\label{polyakovwiegmannnc}
\end{eqnarray}
The action (\ref{todancaction2}) has the left-right local symmetry
\begin{eqnarray}
 e_{\star}^{\phi_k} \rightarrow h_0(z)\star
e_{\star}^{\phi_k(z,\bar{z})}\star \tilde{h}_0(\bar{z}), \quad
\mathrm{for}\quad k=1\dots n, \label{localsymmetrytoda}
\end{eqnarray} where $h_0(z),
\tilde{h}_0(\bar{z})\in G_0$, which is relic of the left-right
local symmetry $g(z,\bar{z}) \rightarrow h(z)\star
g(z,\bar{z})\star \tilde{h}(\bar{z})$ with $h(z),
\tilde{h}(\bar{z})\in G$ of the $WZNW_{\star}$ model, whose
corresponding conserved currents (\ref{chiralcurrents}) close, in
the same way as the ordinary commutative case, a Kac-Moody algebra
\cite{parvizi}. The currents of the abelian Conformal Toda models
(\ref{abeliantoda}) generate a $W$-algebra \cite{bilal}. Since we
have now in our action an infinite number of time derivatives it
is not trivial how to define the conjugate momenta associated to
the fields $\phi_k$ and consequently to define the corresponding
Poisson brackets necessary to study the algebra of currents of the
constrained model.
 Consider the special symmetry subgroup $U_L(1)\times U_R(1)$ where
 $h(z)=U_L(z)=e_{\star}^{i\alpha_1(z)}$ and
 $\tilde{h}(\bar{z})=U_R(\bar{z})=e_{\star}^{i\alpha_2(\bar{z})}$ and $\phi_k \rightarrow i \phi_k$. As
far as
 the global symmetry is concerned there is no difference between
 the point-wise and star product, so only the combination $U_LU_R=e^{i\alpha}$
 is important. Notice that this is a symmetry of the action as well as
of the equations of motion. Let us try to find the corresponding
conserved charge and for this purpose we will localize $\alpha$.
Since the action is invariant for this variation of the field
\begin{eqnarray}
\delta S \equiv -\int d^2x J^{\mu}(\phi) \partial_{\mu}
\alpha(x)=0.
\end{eqnarray}
Integrating by parts
\begin{eqnarray}
 \int d^2x
\partial_{\mu}J^{\mu}\alpha(x)=\int d^2x [f,g]_{\star}=0,
\end{eqnarray}
 for some functions $f,g$. What means that here the notions of conserved current
 and of conserved charged are not well-defined. So it seems that
 Noether's theorem when time is a not-commuting coordinate no longer
applies. For this reason it is not clear what are the conserved
charges asso\-cia\-ted to these symmetries if there is any one.
The study of the symmetries of these theories requires first a nc
analog of Noether theorem that could relate in a possible way
general symmetries to conservation laws.

 So, in this section we have constructed nc analogs of abelian Toda
field theories from a $\star-$zero-curvature condition and in this
way they posses an infinite number of conserved charges.
Associating to these theories an appropriate bicomplex it is
possible to construct by an iterative process an infinite chain of
conserved charges \cite{dimakis}. In the next section, in relation
to the work \cite{dimakis}, we will study the corresponding
bicomplex. Nevertheless the study of the influence of this
infinite chain of conserved charges on the integrability
properties of the theories is still an open question.

\subsection{The nc Abelian Toda field theories as open Toda field
chains}

 The open Toda chains on the usual commutative case are integrable
systems asso\-cia\-ted with finite-dimensional Lie algebras. The
model consists of a one-dimensional chain of $n$ particles with
nonlinear nearest neighbor interaction \cite{Toda}. The
relativistic invariant abelian Conformal Toda field theories
(\ref{abeliantodaeq}) can be called as open Toda field chains
since the equations of motion of these theories reduce to the open
Toda chain equations of motion when the fields do not depend on
$x$. To the Toda field theories it is possible in general to
associate a bicomplex \cite{emilio}, which is a special case of
zero curvature formulation. Noncommutative extensions of
integrable models can be obtained deforming the associated
bicomplex by replacing the ordinary products of functions with the
Moyal product \cite{dimakis}. Following this procedure in
\cite{dimakis} was constructed a nc extension of a Toda field
theory on an open finite one-dimensional lattice up to first order
in $\theta$. It is not difficult to see that expressing the
components of the zero grade element as
$e_{\star}^{\phi_k}=e^{q_k}(1+\theta \tilde{q_k})+O(\theta^2)$ our
system (\ref{todaeqnmotion}) reduces to the proposal of Dimakis-
M\"{u}ller-Hoissen (expressions $(3.10)$ and $(3.11)$ of
\cite{dimakis} mapping $q_k, \tilde{q_k} \rightarrow
-q_k,-\tilde{q_k}$ respectively). At zero order of $\theta$ the
open Toda equations,
 \begin{eqnarray}
(\partial_{t}^2-\partial_{x}^2)q_k=e^{q_{k+1}-q_{k}}-e^{q_{k}-q_{k-1}}
\quad \mathrm{for} \quad k=2...n-1, \\
(\partial_{t}^2-\partial_{x}^2)q_1=e^{q_{2}-q_{1}}, \quad \quad
(\partial_{t}^2-\partial_{x}^2)q_n=-e^{q_{n}-q_{n-1}}, \nonumber
\end{eqnarray}
are obtained. Then at first order of $\theta$ the corresponding
equations are
\begin{eqnarray}
(\partial_{t}^2-\partial_{x}^2)\tilde{q}_1&=&\{\partial_t
q_1,\partial_x q_1
\}+e^{q_{2}-q_{1}}(\tilde{q}_{2}-\tilde{q}_{1}),
\nonumber \\
(\partial_{t}^2-\partial_{x}^2)\tilde{q}_k&=&\{\partial_t
q_k,\partial_x q_k
\}+e^{q_{k+1}-q_{k}}(\tilde{q}_{k+1}-\tilde{q}_{k})-e^{q_{k}-q_{k-1}}(\tilde{q}_{k}-\tilde{q}_{k-1}),
 \nonumber \\
\partial_{t}^2-\partial_{x}^2)\tilde{q}_n&=&\{\partial_t
q_n,\partial_x q_n
\}+e^{q_{n}-q_{n-1}}(\tilde{q}_{n}-\tilde{q}_{n-1}),
\end{eqnarray}
where we have considered $\mu_k=\frac{1}{2}$ for $k=1...n-1$ and
$\{f,g\}=\partial_t f \partial_x g-\partial_x f \partial_t g$.
This fact is not unexpected since the bicomplex equation from
where the Toda chain equations are derived can be written as the
nc Leznov-Saveliev equation (\ref{Lez-Savnc}) as we will see
immediately. Consider the bicomplex equation \cite{dimakis}
 \be
 M_t-M_x=L\star S
-S\star L, \label{bicomplex}
\ee with
\begin{eqnarray}
L=G^{-1}\star S^T \star G, \quad M=2G^{-1}\star\partial G,
\end{eqnarray}
$G$ an invertible $n\times n$ matrix that in \cite{dimakis} is
taken diagonal and $S^{T}$ the transpose of
\begin{eqnarray}
S=\sum_{i=1}^{n-1}E_{i,i+1} \quad \mathrm{with} \quad (E_{i,j})^k
l=\delta^{k}_i\delta_{j,l}.
\end{eqnarray}
Taking into account that the matrix $S$ in this case is nothing
more than the matrix representation of the constant generators of
grade $\pm 1$ (\ref{eps}) and $G$ the matrix representation of the
zero grade group element $B$, the equation (\ref{bicomplex}) can
be written as a nc Leznov-Saveliev equation (\ref{Lez-Savnc})
\footnote{In fact consider in this case $G=B^{-1}$ and
$\epsilon_+=\frac{1}{2}S$.}. In this sense we have the bicomplex
associated to these integrable models what allows to construct
generalized conserved densities. In this specific case the  first
two charges were computed in \cite{dimakis}. The
ge\-ne\-ra\-li\-za\-tion to the closed Toda chain related to loop
algebras and consequently to the abelian affine Toda models is
straightforward.

\subsection{NC Liouville}

Let us concentrate in this part of the section on an specific
example of a Toda field theory, i.e. the Liouville model. This is
an integral and conformal invariant theory that appears in many
applications related to string theory and two-dimensional quantum
gravity. It turns out to be the simplest example of an abelian
Conformal Toda model, connected to $SL(2)$. Following that we can
present a nc extension of this model taking $n=2$ in
(\ref{todaeqnmotion}), say
\begin{eqnarray}
\partial( \bar{\partial}(e_{\star}^{\phi_+})\star
e^{-\phi_+}_{\star})
=\mu^2 e^{\phi_-}_{\star}\star e^{-\phi_+}_{\star}, \nonumber \\
\partial(\bar{\partial}(e_{\star}^{\phi_-})\star
e^{-\phi_-}_{\star})=-\mu^2 e^{\phi_-}_{\star}\star
e^{-\phi_+}_{\star}, \label{eqnliouvillenc2}
\end{eqnarray}
where we have called $\phi_+=\phi_1$, $\phi_-=\phi_2$ and
$\mu_1=\mu$. One can now also compute the sum and difference of
that equations to find
\begin{eqnarray}
\partial(\bar{\partial}(e^{\phi_+}_{\star})\star e^{-\phi_+}
+\bar{\partial}(e^{\phi_-}_{\star})\star e^{-\phi_-})&=&0,
\nonumber
\\
\partial(\bar{\partial}(e^{\phi_+}_{\star})\star e^{-\phi_+}
-\bar{\partial}(e^{\phi_-}_{\star})\star e^{-\phi_-}) &=& 2\mu^2
e_{\star}^{\phi_-}\star e_{\star}^{-\phi_+}.
\label{eqnliouvillenc2.2}
\end{eqnarray}
In this sense (\ref{eqnliouvillenc2}) or (\ref{eqnliouvillenc2.2})
will represent the nc analogs of the Liouville model. Both systems
in the commutative limit will lead to a decoupled model of two
fields
\begin{eqnarray}
\partial \bar{\partial} \varphi_0=0 \quad \mathrm{and} \quad \partial
\bar{\partial} \varphi_1=\mu^2 e^{-2\varphi_1},
\label{liouvillecommutative}
\end{eqnarray}
as we already saw in (\ref{abeliantodaeq}). The first equation in
(\ref{eqnliouvillenc2.2}) transforms to a free field equation for
$\varphi_0$ and the second leads to the usual Liouville equation.

The action corresponding to the nc analog of Liouville model
follows from (\ref{todancaction2}) for $n=2$, namely
\begin{eqnarray}
S(\phi_+,\phi_-)&=&
S_{WZNW_{\star}}(e_{\star}^{\phi_+})+S_{WZNW_{\star}}(e_{\star}^{\phi_-})+\frac{k}{2\pi}\int
d^2z \mu^{2}e_{\star}^{\phi_-}\star e_{\star}^{-\phi_+}.
\label{liouvillencaction2}
\end{eqnarray}

The nonabelian character of the zero grade subgroup allows an
alternative parameterization for the zero grade group element $B$,
say
 \be
B=e^{\varphi_1h}_{\star}\star e^{\varphi_0 I}_{\star} ,
\label{BLiouville}
 \ee
where $h$ is the Cartan and $I$ is the identity generator. We can
keep the same constant generators of grade $\pm 1$,
 \be
 \epsilon_{\pm}=\mu E_{\pm \alpha} \label{epsliouville}.
 \ee
Introduce (\ref{BLiouville}) and (\ref{epsliouville}) in the nc
generalization of Leznov-Saveliev equations (\ref{Lez-Savnc}).
Using algebraic properties it is very simple to see that
 \be
  B \star \epsilon_- B^{-1}=
e^{-2\varphi_1}_{\star}\epsilon_-.
 \ee
For computing the equations of motion, notice that the matrix
representation (\ref{representation}) reduces to the usual
representation
\begin{eqnarray}
E_{\alpha}=\left( \begin{array}{cc}
                          0 & 1 \\
                          0 & 0
                     \end{array} \right),       \quad
E_{-\alpha}=\left( \begin{array}{cc}
                           0 & 0 \\
                           1 & 0
                           \end{array}
                           \right),
                           \quad
h=\left( \begin{array}{cc}
                           1 & 0 \\
                           0 & -1
                           \end{array}
                           \right), \label{pauli}
\end{eqnarray}
for ${\cal SL}(2)$. In this way one find that the equations that
define the nc extension of Liouville model in this alternative
parameterization are
\begin{eqnarray}
\partial( \bar{\partial}(e_{\star}^{\varphi_1}\star
e^{\varphi_0}_{\star}) \star e^{-\varphi_0}_{\star}\star e^{
-\varphi_1}_{\star})
&=&\mu^2 e^{-2\varphi_1}_{\star}, \nonumber \\
\partial(\bar{\partial}(e_{\star}^{-\varphi_1}\star
e^{\varphi_0}_{\star})\star e^{-\varphi_0}_{\star}\star
e^{\varphi_1}_{\star})& =&-\mu^2 e^{-2\varphi_1}_{\star}.
\label{eqnliouvillenc1}
\end{eqnarray}
One can also combine the previous equations to find the system of
coupled second order equations,
\begin{eqnarray}
\partial(\bar{\partial}(e^{\varphi_1}_{\star}\star
e^{\varphi_0}_{\star})\star e_{\star}^{-\varphi_0}\star
e^{-\varphi_1}_{\star} +\bar{\partial}(e_{\star}^{-\varphi_1}\star
e^{\varphi_0}_{\star})\star
e_{\star}^{-\varphi_0}\star e^{\varphi_1}_{\star})&=& 0,  \nonumber \\
\partial(\bar{\partial}(e^{\varphi_1}_{\star}\star
e^{\varphi_0}_{\star})\star e_{\star}^{-\varphi_0}\star
e^{-\varphi_1}_{\star} -\bar{\partial}(e_{\star}^{-\varphi_1}\star
e^{\varphi_0}_{\star})\star e_{\star}^{-\varphi_0}\star
e^{\varphi_1}_{\star}) &=&2\mu^2 e_{\star}^{-2\varphi_1},
\label{eqnliouville1.1}
\end{eqnarray}
which are more suitable for applying the commutative limit. When
$\theta \rightarrow 0$ that system easily reduced to
(\ref{liouvillecommutative}).

 One can write an action for the nc Liouville model
(\ref{eqnliouvillenc1}) using the general expression
(\ref{effectiveactionnc}) and making use of the nc generalization
of the Polyakov-Wiegmann identity (\ref{polyakovwiegmannnc}). For
this purpose, introduce (\ref{BLiouville}) and
(\ref{epsliouville}) in (\ref{effectiveactionnc}) to obtain the
corresponding action of nc Liouville (\ref{eqnliouvillenc1})
\begin{eqnarray}
S(\varphi_1,\varphi_0)&=&
2S_{PC_{\star}}(e_{\star}^{\varphi_1})+2S_{WZNW_{\star}}(e_{\star}^{\varphi_0})+\frac{k}{2\pi}\int
d^2z \mu^{2}e_{\star}^{-2\varphi_1} - \nonumber \\ & &-
\frac{k}{2\pi}\int d^2z ( \bar{\partial}e_{\star}^{\varphi_0}\star
e_{\star}^{-\varphi_0} \star( e_{\star}^{-\varphi_1}\star
\partial e_{\star}^{\varphi_1}+ e_{\star}^{\varphi_1} \star
\partial e_{\star}^{-\varphi_1})), \label{liouvillencaction1}
\end{eqnarray}
where using the notation of \cite{lechtenfeld} we have defined
\begin{eqnarray}
S_{PC_{\star}}(g)=-\frac{k}{4\pi}\int_{\Sigma} d^2z Tr
(g^{-1}\star
\partial{g} \star g^{-1} \star \bar{\partial}g).
\end{eqnarray}
Looking at the action (\ref{liouvillencaction1}) it is noticeable
the presence of the topological term of $WZNW_{\star}$, in this
case shifted to the $\varphi_0$ field. This situation contrasts
with the ordinary commutative case, where for Liouville and more
generally for any abelian subspace this term equals zero.

The parameterizations (\ref{BLiouville}) and
(\ref{liouvillecommutative}) for the element $B$ belonging to the
zero grade subgroup in the usual commutative case are identical,
but in a nc space-time they lead to equivalent models. Taking
\begin{eqnarray}
e^{\varphi_1}_{\star}\star e^{\varphi_0}_{\star}\rightarrow
e^{\phi_+}_{\star} \quad \mathrm{and} \quad
e^{-\varphi_1}_{\star}\star e^{\varphi_0}_{\star}\rightarrow
e^{\phi_-}_{\star},
\end{eqnarray}
(\ref{eqnliouvillenc1}) can be transformed in
(\ref{eqnliouvillenc2}). It is interesting to note that the usual
Liouville singular solution \cite{balog}
\be \varphi_1=\ln \cos
(\alpha z -\beta \bar{z})\ee is also solution of the systems
(\ref{eqnliouvillenc2}) and (\ref{eqnliouvillenc1}) as far as
$\varphi_0=\alpha z -\beta \bar{z}$ and the constant parameters
$\alpha,\beta$ satisfy $\mu^2=\alpha \beta$. This is a consequence
of the fact that for this particular dependence of the fields
$\varphi_1, \varphi_0$ on the variables $z,\bar{z}$ the
star-product reduce to the usual product. So in this case
$\varphi_0$ decouples and the nc model is reduced to the usual
Liouville theory plus the equation for a free field. Other
solutions of this model will be discussed in \cite{sdchsyo}.

\subsection{NC Liouville from nc self-dual Yang-Mills}

It has been known for a long time that many two-dimensional
integrable models can be obtained from the Yang-Mills self-duality
equations in four dimensions by reductions \cite{ward}. Recently a
nc extension of the self-dual Yang-Mills (NCSDYM) equations in the
Yang formulation \cite{takasaki},
 \be
\partial_{y}(\partial_{\bar{y}} J \star
J^{-1})+\partial_{z}(\partial_{\bar{z}} J \star J^{-1})=0,
\label{yang}
 \ee
has been proposed. In (\ref{yang}) $y, \bar{y}, z, \bar{z}$ are
complex independent variables $y=x_1+ix_2$, $\bar{y}=x_1-ix_2$,
$z=x_3-ix_4$, $\bar{z}=x_3+ix_4$ that do not commute.
 Moreover, from (\ref{yang}) through a
dimensional reduction process some nc extensions of
two-dimensional integrable models have been obtained
\cite{integrablenc2, GP2, lechtenfeld}. In this part of the
section we will show how the systems of equations
(\ref{eqnliouvillenc2}) and (\ref{eqnliouvillenc1}) that define
the nc extensions of Liouville model can be also derived from
(\ref{yang}) through a suitable dimensional reduction. For this
purpose consider \footnote{This is a nc extension of the $J$
decomposition proposed in \cite{mo} for Liouville. } \be
J=e_{\star}^{wE_{\alpha}}\star B \star e_{\star}^{-wE_{-\alpha}},
\ee where $w=\mu(y+\bar{y})$ with $\mu$ a constant parameter and
$B$ a zero grade group element in this case of $GL(2,
\mathbb{C})$, whose fields depend only on the variables $z,
\bar{z}$.

Let us denote $M=e_{\star}^{w E_{\alpha}}$.  It is straightforward
to see that (\ref{yang}) reduces to
\begin{eqnarray}
M \star \{\partial_{z}(\partial_{\bar{z}} B \star B^{-1})-\mu^2(
E_{\alpha}B E_{-\alpha}\star B^{-1}-B E_{-\alpha}\star
B^{-1}E_{\alpha} )\}\star M^{-1}=0.
\end{eqnarray}
Remembering that for Liouville $\epsilon_{\pm}=\mu E_{\pm
\alpha}$, the previous equation will render \be
\partial_{z}(\partial_{\bar{z}} B \star B^{-1})-[\epsilon_+ ,B
\epsilon_- \star B^{-1}]_{\star}=0.
 \ee
Taking $x_3=t$ and $x_4=ix$ this is the nc Leznov-Saveliev
equation (\ref{Lez-Savnc}) from where (\ref{eqnliouvillenc1}) and
(\ref{eqnliouvillenc2}) were derived. In this sense, we have shown
that the nc analogs of Liouville theory (\ref{eqnliouvillenc2})
and (\ref{eqnliouvillenc1}) can be obtained through an appropriate
reduction of the nc self-dual Yang-Mills system in the same way as
in the ordinary commutative case. More general, in this work we
will see how the nc Leznov-Saveliev equations (\ref{Lez-Savnc})
can be obtained through a dimensional reduction process from the
nc (anti-)self-dual Yang-Mill equations via the nc self-dual
Chern-Simons system.

\subsection{Other proposals}

As was mentioned in the previous work \cite{us} the nc extension
of a field theory is not unique. In this sense a different nc
proposal for the Liouville model was presented in \cite{zuevsky}
\be \bar{\partial} \left((e_{\star}^{\beta \phi})^{-1}_{\star L}
\star
\partial e_{\star}^{\beta \phi} \right)=e_{\star}^{\beta \phi},
 \ee
where $(e_{\star}^{\beta \phi})^{-1}_{\star L}$ is denoted as the
left inverse function of $e_{\star}^{\beta \phi}$ with respect to
the $\star$-product and $\beta$ a constant. This equation is
obtained from a zero-curvature condition on the basis of a
generalization of Saveliev-Vershik continual Lie algebras
\cite{save-ver}. Note that in this case the model is defined
through only one equation. Not presented in \cite{zuevsky}, but it
can be obtained as the Euler-Lagrange equation of motion of the
action
\begin{eqnarray}
S_{L-1}=S_{WZNW_{\star}}(e_{\star}^{\beta\phi})+\frac{k}{2\pi}\int
d^2z e_{\star}^{\beta \phi}, \label{zuevski}
\end{eqnarray}
which can be thought as the naive nc generalization of the $SL(2)$
Toda theory at the level of the action just generalizing the
derivative term. In the case of sinh/sine Gordon model studied in
\cite{us} this procedure leads to a {\it non}-integrable
deformation since the amplitude for the scattering $2 \rightarrow
4$ process it is non-zero. This fact is in some sense explained by
the difficulty to find a zero curvature representation for this
deformation without the inclusion of extra conditions. In the case
of (\ref{zuevski}) we failed to find a zero-curvature
representation in terms of the Lie algebra $SL(2)$. But this does
not mean that the model is not integrable. In order to check the
integrability properties of (\ref{zuevski}) could be interesting
to study the properties of the corresponding S-matrix. By the
other side a different nc extension
\begin{eqnarray}
\partial( \bar{\partial}(e_{\star}^{i\varphi_1}
) \star e^{ -i\varphi_1}_{\star})
&=&\mu^2 e^{-2i\varphi_1}_{\star}, \nonumber \\
\partial(\bar{\partial}(e_{\star}^{-i\varphi_1} )\star
 e^{i\varphi_1}_{\star})& =&-\mu^2
e^{-2i\varphi_1}_{\star}, \label{eqnliouvillenc3}
\end{eqnarray}
could in principle be constructed from the $\star$-zero-curvature
condition (\ref{starzc}) for the algebra $SL(2)$, excluding the
direction of the identity generator of the Cartan subalgebra (in
this specific case we have considered that $\varphi_1 \rightarrow
i\varphi_1$). In order to find an action for
(\ref{eqnliouvillenc2}) or equivalently for the system
\begin{eqnarray}
\partial( \bar{\partial}(e_{\star}^{i\varphi_1}
) \star e^{
-i\varphi_1}_{\star})+\partial(\bar{\partial}(e_{\star}^{-i\varphi_1}
)\star
 e^{i\varphi_1}_{\star})
&=& 0,\nonumber \\
\partial( \bar{\partial}(e_{\star}^{i\varphi_1}
) \star e^{
-i\varphi_1}_{\star})-\partial(\bar{\partial}(e_{\star}^{-i\varphi_1}
)\star
 e^{i\varphi_1}_{\star})& =&2\mu^2
e^{-2i\varphi_1}_{\star}, \label{eqnliouvillenc3.1}
\end{eqnarray}
it can be followed the procedure in \cite{GP2}. So one can
consider that if $\varphi_1$ is a complex field then $e_{\star}^{i
\varphi_1^{\dagger}}$ satisfies the equations
\begin{eqnarray}
\partial( \bar{\partial}(e_{\star}^{i\varphi_1^{\dagger}}
) \star e^{ -i\varphi_1^{\dagger}}_{\star})
&=&-\mu^2 e^{2i\varphi_1^{\dagger}}_{\star}, \nonumber \\
\partial(\bar{\partial}(e_{\star}^{-i\varphi_1^{\dagger}} )\star
 e^{i\varphi_1^{\dagger}}_{\star})& =&\mu^2
e^{2i\varphi_1^{\dagger}}_{\star}, \label{eqnliouvillenc4}
\end{eqnarray}
 obtained by hermitian conjugation of (\ref{eqnliouvillenc3}). An action for this model could be
\begin{eqnarray}
S_{L-2}=S_{WZNW_{\star}}(e_{\star}^{i\varphi_1})+S_{SWZN_{\star}}(e_{\star}^{-i\varphi_1^{\dagger}})+\frac{k}{2\pi}\int
d^2z(e_{\star}^{i2\varphi_1^{\dagger}}+e_{\star}^{-i2\varphi_1}) ,
\end{eqnarray}
if as independent equations are considered the first in
(\ref{eqnliouvillenc3}) and the second in (\ref{eqnliouvillenc4}).
In \cite{kimeon} was proposed another nc extension of the
Liouville model starting from the nc self-dual Chern-Simons
equations. We will discuss about this point later in this paper.

\section{NC $\widetilde{GL}(n)$ abelian affine Toda field theories}

Among Toda theories the affine models are quite special due,
essentially, to the presence of soliton type solutions. Looking at
possible applications in particle physics and in condensed matter,
in this section we will extend these theories to the nc scenario.

Consider now the nc generalization of the two-loop $WZNW$ model
\footnote{See \cite{aratyn} for the ordinary commutative case.},
which is formally described by the same action (\ref{wznwnc}), but
in this case ${\cal G}$ is an infinite-dimensional algebra (see
the appendix). In the same way $N$ and $M$ are
infinite-dimensional subalgebras. The zero grade subgroup $G_0$ is
however chosen to be finite-dimensional. Imposing appropriate
constraints on the chiral currents (\ref{HRconstraints}) in such a
way that now the constant generators of grade $\pm 1$ include
extra affine generators,
\begin{eqnarray}
\epsilon_{\pm}=\sum_{i=1}^{n-1}\mu_iE_{\pm \alpha_i}^{(0)}+m_0
E^{(\pm1)}_{\mp\psi}, \label{epsaffine}
\end{eqnarray}
where $\psi$ is the highest root of ${\cal G}={\cal SL}(n)$ and
$m_0$, $\mu_i$ with $i=1\dots n-1$ constant parameters, the
equations of motion (\ref{eqnwznwnc}) reduce to the affine version
of the nc Leznov-Saveliev equations (\ref{Lez-Savnc}). For this
reason we are going to use them again as starting point in order
to define the nc extensions of $\widetilde{GL}(n)$ abelian affine
Toda theories.

The grading operator $Q$ in the principal gradation for the affine
algebra $\widetilde{{\cal GL}}(n)$ is taken as
 \be
 Q=\sum_{i=1}^{n-1}\frac{2\lambda_i \cdot
H^{(0)}}{\alpha_i^2}+n d,
 \ee
where $d$ is the derivation generator and its coefficient is
chosen such that this gradation ensures that the zero grade
subspace ${\cal G}_0$ coincides with its counterpart on the
corresponding Lie algebra ${\cal SL}(n)$, apart from the generator
$d$. The abelian subalgebra of grade zero is in this case ${\cal
G}_0=\{I, h_1^{(0)},h_2^{(0)},\dots h_{n-1}^{(0)}, d \}$ and for
this reason the zero grade group element $B$ is parameterized as
the $\star$-exponentiation of these generators as in (\ref{B})
\footnote{The field associated to the derivation generator $d$ has
been set to zero as is usually done for the affine theories.}.
Working in the $n \times n$ representation,
\begin{eqnarray}
(h_i)^{(0)}_{\mu\nu}=\delta_{\mu\nu}(\delta_{i,\mu}-\delta_{i+1,\mu}),
\quad (E^{(\pm1)}_{\alpha_i})_{\mu \nu}=
\lambda^{\pm1}\delta_{\mu,i}\delta_{\nu,i+1}, \quad
(E^{(\pm1)}_{-\alpha_i})_{\mu \nu}=
\lambda^{\pm1}\delta_{\nu,i}\delta_{\mu,i+1}, \nonumber
\end{eqnarray}
where $\lambda$ is the spectral parameter and using the variables
(\ref{newvariables}), the components of the gauge potentials
(\ref{potentialnc}) read
\begin{eqnarray}
\bar{A}_{ij}&=&\bar{\partial}(e_{\star}^{\phi_i})\star
e_{\star}^{-\phi_i}\delta_{ij}+\mu_i \delta_{i+1,j}+\lambda m_0
\delta_{i,n}\delta_{j,1}, \nonumber \\
A_{ij}&=&-\mu_ie_{\star}^{\phi_{i+1}}\star
e^{-\phi_i}_{\star}\delta_{i,j+1}-\frac{m_0}{\lambda}e_{\star}^{\phi_1}\star
e_{\star}^{-\phi_n}\delta_{i,1}\delta_{j,n}.
\label{todapotentialsaffinenc}
\end{eqnarray}
Introducing the potentials (\ref{todapotentialsaffinenc}) in the
${\star}$-zero-curvature condition (\ref{starzc}), the n-coupled
equations of motion,
\begin{eqnarray}
\partial(\bar{\partial}( e_{\star}^{\phi_{k}})\star
e_{\star}^{-\phi_{k}})&=& \label{todaeqnaffinenc} \\
&&\mu_{k}^2e_{\star}^{\phi_{k+1}}\star e_{\star}^{-\phi_{k}}
-\mu_{k-1}^2 e_{\star}^{\phi_{k}}\star
e_{\star}^{-\phi_{k-1}}+m^2_0(\delta_{n,k}-\delta_{1,k})
e_{\star}^{\phi_1}\star e_{\star}^{-\phi_n}, \nonumber
\end{eqnarray}
are found. Note that in the previous expression $\mu_0=\mu_n=0$
and $k=1\dots n$. Applying the commutative limit, the system of
equations,
\begin{eqnarray}
\bar{\partial}\partial
\varphi_i&=&\mu_i^2e^{-k_{ij}\varphi_j}-m_0^2e^{k_{\psi
j}\varphi_j}, \nonumber \\
n \bar{\partial} \partial \varphi_0 & = & 0,
 \label{affinetodaeqm}
\end{eqnarray}
for the original variables (\ref{newvariables}) is obtained, where
$k_{\psi j}=\frac{2 \psi \cdot \alpha_j}{\alpha_j^2}$ is the
extended Cartan matrix. These are the equations of motion of the
affine Toda model plus an additional equation for a free scalar
field. The action from where the nc $\widetilde{GL}(n)$ affine
equations (\ref{todaeqnaffinenc}) can be derived reads
\begin{eqnarray}
S(\phi_1,\dots,\phi_n)&=& \label{todancactionaffine} \\
&&
\sum_{k=1}^{n}S_{WZNW_{\star}}(e_{\star}^{\phi_k})+\frac{k}{2\pi}\int
d^2z \left(\sum_{k=1}^{n-1}\mu^2_k e_{\star}^{\phi_{k+1}}\star
e_{\star}^{-\phi_{k}}+m^2_0e_{\star}^{\phi_1}\star
e_{\star}^{-\phi_n}\right), \nonumber
\end{eqnarray}
which in the commutative limit reduces to
\begin{eqnarray}
S(\varphi_1,\dots,\varphi_{n-1},\varphi_0)&=&S_{AT}(\varphi_1,\dots,\varphi_{n-1})+
nS_0(\varphi_0),
\end{eqnarray}
with
\begin{eqnarray}
S_{AT}(\varphi_1,\dots,\varphi_{n-1})&=& -\frac{k}{4\pi}\int d^2z
\left( k_{ij}\partial \varphi_i \bar{\partial} \varphi_j-\left(2
\sum_{i=1}^{n-1}\mu_i^2 e^{-k_{ij}\varphi_j}+2m_0^2
e^{k_{\psi_j}\varphi_j}\right)\right), \nonumber \\
\label{affinetoda}
\end{eqnarray}
i.e. the usual affine Toda action plus the action for a free
field. The complex $\widetilde{SL}(n)$ nc abelian affine Toda
field theory (\ref{todancactionaffine}) ($\phi_k \rightarrow i
\phi_k$) satisfies the usual abelian affine one-soliton solution
\cite{olive1,hollowood}
\begin{eqnarray}
i\varphi_k&=& \ln\left(\frac{1+e^{\sigma x-\lambda
t+\xi+\frac{2\pi ia}{n}k}}{1+e^{\sigma x-\lambda t+\xi+2\pi
ia}}\right), \label{affine solitons}
\end{eqnarray}
where we have considered that $\mu_k=m_0=m$ for $k=1$ to $n-1$,
$\sigma$ and $\lambda$ are real parameters satisfying
$\sigma^2-\lambda^2=16m^2\sin^2\frac{\pi a}{n}$, $a$ is an integer
in the set $\{1,2,\dots n-1 \}$ and $\xi$ is an arbitrary complex
parameter. This is true since we have considered for the extra
field $\varphi_0$ the solution $\varphi_0=\sigma x-\lambda t$. In
this case due to the particular dependence of $\phi_k$ on the
variables $x,t$ the star product of fields become the usual
product. The solitons (\ref{affine solitons}) are kinks, whose
center of mass is at $\sigma^{-1}(\lambda t -\mathrm{Re} \xi)$,
they move with velocity $\frac{\lambda}{\sigma}$ and have the
characteristic size $\sigma^{-1}$. To the soliton type solutions
of (\ref{todancactionaffine}) we can associate a topological
charge \be Q_k=\frac{1}{2 \pi} \int_{-\infty}^{\infty} dx
\frac{\partial \varphi_k}{\partial x},  \quad \mathrm{for} \quad
k=1\dots n-1, \ee which for the specific one-soliton solutions
(\ref{affine solitons}) will depend on the parameters $a$ and
$\mathrm{Im} \xi$. This charge is associated to the discrete
symmetry of the model (\ref{todancactionaffine}) $i\phi_k
\rightarrow i\phi_k +2i\pi l$ with $l$ an integer number and in
fact due to the infinite generalizations possible for the
derivative terms there is an ambiguity in its definition.
 The study of multisolitons solutions of these
theories could be possible achieved by a nc version of the
dressing method \cite{dressing}. We expect that at this level the
solution for $\varphi_0$ would not be trivial. In
\cite{lechtenfeld} was outlined the construction of nc
multisolitons through this method for the specific case of the
sine-Gordon model. Nevertheless the explicit construction of
multisoliton solutions for this theory has not been performed yet.

 The action (\ref{todancactionaffine}) also has the left-right local
symmetry (\ref{localsymmetrytoda}) and the global $U_L(1)\times
U_R(1)$ $e_{\star}^{\phi_k} \rightarrow
e^{i\alpha_1}e_{\star}^{\phi_k} e^{i\alpha_2}$. This point seems
interesting since in case of this symmetry being associated to
some conserved charge, the soliton type solutions of the nc
abelian affine Toda field theories could possible carry electric
charge, what it does occur in the usual case.

\subsection{NC sinh/sine-Gordon}

The sinh/sine-Gordon model, connected with the $\widetilde{{\cal
SL}}(2)$ loop algebra is the simplest example of an abelian affine
Toda theory. In this sense a nc sinh/sine-Gordon extension has
been already contemplated in (\ref{todaeqnaffinenc}). Considering
$n=2$, $\phi_1=\phi_+$, $\phi_2=\phi_-$ and $m_0=\mu_1=\mu$
(\ref{todaeqnaffinenc}) reduced to
\begin{eqnarray}
\partial(\bar{\partial}(e^{\phi_+}_{\star})\star
e^{-\phi_+}_{\star}) &=&\mu^2(e_{\star}^{\phi_-}\star
e_{\star}^{-\phi_+}-e_{\star}^{\phi_+}\star e_{\star}^{-\phi_-}), \nonumber \\
\partial(\bar{\partial}(e^{\phi_-}_{\star})\star
e^{-\phi_-}_{\star})&=&\mu^2(-e_{\star}^{\phi_-}\star
e_{\star}^{-\phi_+}+e_{\star}^{\phi_+}\star e_{\star}^{-\phi_-}).
\label{eqnsinegordonnc2}
\end{eqnarray}
Computing the sum and difference of the previous equations
\begin{eqnarray}
\partial(\bar{\partial}(e^{\phi_+}_{\star})\star e^{-\phi_+}_{\star}
+\bar{\partial}(e^{\phi_-}_{\star})\star e^{-\phi_-}_{\star})&=&0,
\nonumber \\
\partial(\bar{\partial}(e^{\phi_+}_{\star})\star e^{-\phi_+}_{\star}
-\bar{\partial}(e^{\phi_-}_{\star})\star e^{-\phi_-}_{\star}) &=&
2\mu^2(e_{\star}^{\phi_-}\star
e_{\star}^{-\phi_+}-e_{\star}^{\phi_+}\star e_{\star}^{-\phi_-}),
\label{eqnsinegordonnc2.2}
\end{eqnarray}
an equivalent version of nc sinh-Gordon is obtained. The complex
version of this system have been already presented in the
literature \cite{lechtenfeld}. There (\cite{lechtenfeld}) the nc
extension of sine-Gordon model was constructed through a
dimensional reduction process, star\-ting from the linear system
of the nc extension of the $2+1$ sigma model, which in
\cite{popov2} was shown to be integrable. Here we have seen how
this model can be also obtained directly in two-dimensions from
the nc extension of the Leznov-Saveliev equations
(\ref{Lez-Savnc}). The first of these two equations belongs also
to the system (\ref{eqnliouvillenc2.2}) that defines the nc
Liouville generalization in this parameterization. In the
commutative limit, this equation produces a free field equation
for $\varphi_0$ and the second one reduces to the usual
sinh-Gordon equation
 \be
\partial \bar{\partial} \varphi_1+2 \mu^2 \sinh 2\varphi_1=0.
\label{sine-Gordon}
 \ee
 In \cite{harold}, the version (\ref{eqnsinegordonnc2}) of
nc sine-Gordon was also derived through a reduction process from
the nc affine Toda model coupled to matter fields.

The action (\ref{todancactionaffine}) corresponding to this nc
generalization of the sinh-Gordon model reads
\begin{eqnarray}
S(\phi_+,\phi_-)=
S_{WZNW_{\star}}(e_{\star}^{\phi_+})+S_{WZNW_{\star}}(e_{\star}^{\phi_-})&+&
\\& + & \frac{k}{2\pi}\int
 d^2z \mu^{2}(e_{\star}^{\phi_-}\star
e_{\star}^{-\phi_+}+e_{\star}^{\phi_+}\star e_{\star}^{-\phi_-}).
\nonumber \label{sinegordonncaction2}
\end{eqnarray}
One can define, in the same way, an equivalent sinh/sine-Gordon
model using the alternative parameterization (\ref{BLiouville}),
i.e.
\begin{eqnarray}
\partial(\bar{\partial}(e^{\varphi_1}_{\star}\star
e^{\varphi_0}_{\star})\star e_{\star}^{-\varphi_0}\star
e^{-\varphi_1}_{\star})
&=&\mu^2(e_{\star}^{-2\varphi_1}-e_{\star}^{2\varphi_1}),  \nonumber \\
\partial(\bar{\partial}(e_{\star}^{-\varphi_1}\star
e^{\varphi_0}_{\star})\star e_{\star}^{-\varphi_0}\star
e^{\varphi_1}_{\star})
&=&-\mu^2(e_{\star}^{-2\varphi_1}-e_{\star}^{2\varphi_1}).
\label{eqnsinegordonnc1}
\end{eqnarray}
Let us now also present the sum and difference of the previous
equations
\begin{eqnarray}
\partial(\bar{\partial}(e^{\varphi_1}_{\star}\star
e^{\varphi_0}_{\star})\star e_{\star}^{-\varphi_0}\star
e^{-\varphi_1} +\bar{\partial}(e_{\star}^{-\varphi_1}\star
e^{\varphi_0}_{\star})\star e_{\star}^{-\varphi_0}\star
e^{\varphi_1}_{\star})&=& 0,
\label{eqnsinegordonnc1.1} \\
\partial(\bar{\partial}(e^{\varphi_1}_{\star}\star
e^{\varphi_0}_{\star})\star e_{\star}^{-\varphi_0}\star
e^{-\varphi_1} -\bar{\partial}(e_{\star}^{-\varphi_1}\star
e^{\varphi_0}_{\star})\star e_{\star}^{-\varphi_0}\star
e^{\varphi_1}_{\star})
&=&2\mu^2(e_{\star}^{-2\varphi_1}-e_{\star}^{2\varphi_1}).
\nonumber
\end{eqnarray}
The complex version of this system ($\varphi_1 \rightarrow i
\varphi_1, \quad \varphi_0 \rightarrow i \varphi_0$) also
reproduce another suggestion for nc sine-Gordon presented in
\cite{lechtenfeld}. Notice that the first equation of this system
belongs to the nc generalization of the Liouville model
(\ref{eqnliouville1.1}) too and it in the commutative limit
$\theta \rightarrow 0$ becomes a free field equation for
$\varphi_0$. The second equation of (\ref{eqnsinegordonnc1.1}) in
the same limit produces the usual sinh-Gordon equation
(\ref{sine-Gordon}).

The action for (\ref{eqnsinegordonnc1})
 can be obtained from
(\ref{effectiveactionnc}) with $B$ as in (\ref{BLiouville}) and
the constant generators $\epsilon_{\pm}$ as in (\ref{epsaffine}),
namely
\begin{eqnarray}
S(\varphi_1,\varphi_0)&=&
2S_{PC_{\star}}(e_{\star}^{\varphi_1})+2S_{WZNW_{\star}}(e_{\star}^{\varphi_0})+\frac{k}{2\pi}\int
d^2z \mu^{2}(e_{\star}^{-2\varphi_1}+e_{\star}^{2\varphi_1})-
\nonumber \\ & & -\frac{k}{2\pi}\int d^2z
(\bar{\partial}e_{\star}^{\varphi_0}\star e_{\star}^{-\varphi_0}
\star( e_{\star}^{-\varphi_1}\star
\partial e_{\star}^{\varphi_1}+ e_{\star}^{\varphi_1} \star
\partial e_{\star}^{-\varphi_1})). \label{sinegordonncaction1}
\end{eqnarray}
As a particular case of (\ref{affine solitons}) for $n=2$, the
one-soliton
\begin{eqnarray}
i\varphi_1&=&2\tan^{-1}(e^{\sigma x-\lambda t+\xi})
\end{eqnarray}
is a solution of (\ref{eqnsinegordonnc1}) and
({\ref{eqnsinegordonnc2}}) with $\sigma^2-\lambda^2=16m^2$. To
compute the scattering amplitudes even for this theory, the
simplest of all the affine Toda theories is not so simple. For
this reason in \cite{lechtenfeld} the scattering amplitudes were
calculated only up to tree level and only for some particle
dispersion processes. Apparently no particle production seems to
occur what could lead to a factorized S-matrix. In this sense, the
nc sine-Gordon model as defined in (\ref{sinegordonncaction1})
seems to retain the integrability properties of the original
model. We expect that the other nc affine Toda theories (those
with $n > 2$) will behave in the same way, although the
corresponding S-matrices have not been investigated so far.

\subsection{About previous suggestions}

{\bf The nc sine-Gordon by Grisaru-Penati:}
\begin{eqnarray}
\partial(\bar{\partial}(e^{\frac{i}{2}\phi}_{\star})\star e^{-\frac{i}{2}\phi}_{\star}
+\bar{\partial}(e^{-\frac{i}{2}\phi}_{\star})\star
e^{\frac{i}{2}\phi}_{\star})&=&0,
\nonumber \\
\partial(\bar{\partial}(e^{\frac{i}{2}\phi}_{\star})\star e^{-\frac{i}{2}\phi}_{\star}
-\bar{\partial}(e^{-\frac{i}{2}\phi}_{\star})\star
e^{\frac{i}{2}\phi}_{\star}) &=& 4\mu^2 \sin_{\star} \phi,
\label{eqnsinegordonncgpandus}
\end{eqnarray}
was presented in \cite{GP1, us}. The system
(\ref{eqnsinegordonncgpandus}) was derived in \cite{us} starting
from the nc extension of the Leznov-Saveliev equations
(\ref{Lez-Savnc}), but excluding of the zero grade subspace the
direction of the identity generator. So, essentially that means
that we were considering the ${\cal SL}(2)$ algebra and the
$GL(2)$ group (as the $SL(2)$ group is not closed on the nc
setting). From our point of view this fact spoils the
integrability properties of the theory. That is why the tree level
scattering amplitude of (\ref{eqnsinegordonncgpandus}) suffers
from acasual behavior and production of particles occurs
\cite{GP2}. By the other side, we expect that the models deformed,
establishing the appropriate {\it algebra-group} relation, could
preserve the integrability properties of their commutative
counterparts.

{\bf The nc sine-Gordon by Zuevsky:}
\begin{eqnarray}
\bar{\partial} \left((e_{\star}^{\beta \phi})^{-1}_{\star L} \star
\partial e_{\star}^{\beta \phi} \right)=\frac{1}{2}(e_{\star}^{\beta
\phi}-e_{\star}^{-\beta \phi}),
\end{eqnarray}
presented in \cite{zuevsky} is obtained from a generalized zero
curvature condition based on continual algebras. In fact this is
the corresponding equation of motion of a deformation of a
sine-Gordon like action done substituting the product of fields by
the star product and generalizing the derivative terms as
$\partial \phi \rightarrow e_{\star}^{-\phi} \star
\partial e_{\star}^{\phi}$. So taking $\beta=1$ it can be derived from the action
principle
\begin{eqnarray}
S_{sG-1}=S_{WZNW_{\star}}(e_{\star}^{\phi})+\frac{k}{4\pi}\int
d^2z \mu^{2}(e_{\star}^{\phi}-e_{\star}^{-\phi}). \label{zuevski1}
\end{eqnarray}
It happened to be difficult to find for it a zero curvature
representation in terms of the generators of the
$\widetilde{SL}(2)$ algebra as in the usual case, so it could be
interesting to investigate the properties of its S-matrix in order
to test its integrability.

\subsection{NC sinh/sine-Gordon from nc self-dual Yang-Mills}

 The nc sinh-Gordon models (\ref{eqnsinegordonnc2}) and
(\ref{eqnsinegordonnc1}) can be derived from nc self-dual
Yang-Mills (\ref{yang}) considering
 \be
 J=e_{\star}^{\mu y(E_{\alpha}+\lambda E_{-\alpha})}\star B
\star e_{\star}^{-\mu \bar{y}(E_{-\alpha}+\frac{1}{\lambda}
E_{\alpha})},
 \ee
where $\lambda$ represents the spectral parameter. It is not
difficult to see that (\ref{yang}) with $x_3=t$ and $x_4=ix$, in
this case reduces to
\begin{eqnarray}
M \star \{ \partial (\bar{\partial} B \star B^{-1})- \epsilon_+ B
\epsilon_-\star B^{-1} + B \epsilon_-\star B^{-1}\epsilon_+
\}\star M^{-1}=0, \label{a}
\end{eqnarray}
where $M=e_{\star}^{\mu y(E_{\alpha}+\lambda E_{-\alpha})}$ and
where we have introduced the constant generators
$\epsilon_{\pm}=\mu (E_{\pm \alpha}+E_{\mp\alpha}^{(\pm 1)})$ for
$\widetilde{{\cal SL}}(2)$. In this way (\ref{a}) represents the
nc sine-Gordon extensions (\ref{eqnsinegordonnc2}) and
(\ref{eqnsinegordonnc1}) with the appropriate zero grade element
$B$.

Before we concluded this section let us remark that in the
construction of the nc gene\-ra\-li\-za\-tions of abelian and
abelian affine Toda theories we have chosen only one type of
parameterization for the element $B$ of the zero grade subgroup.
Whereas it is possible to use the alternative parameterization \be
B=(\prod_{i=1}^{n-1} e^{\varphi_i h^{(0)}_i}_{\star})\star
e_{\star}^{\varphi_0 I},
 \ee
which will lead to generalizations of (\ref{eqnliouvillenc1}) and
(\ref{eqnsinegordonnc1}) for $n>2$.

\section{NC Toda field theories from nc self-dual Chern-Simons}

In \cite{kimeon} was proposed a generalization to the nc plane of
the Toda and affine Toda mo\-dels considering as starting point a
nc extension of the Dunne-Jackiw-Pi-Trugenber (DJPT) \cite{dunne1}
model of a $U(N)$ Chern-Simons gauge theory  coupled to a
nonrelativistic complex bosonic matter field on the adjoint
representation. The lowest energy solutions of this model satisfy
a nc extension of the self-dual Chern-Simons equations from where
through a proposed ansatz, the nc generalizations of Toda and
affine Toda theories were constructed \cite{kimeon}. Although in
the commutative case this procedure will lead to the well known
second order differential equations of the Conformal Toda or
affine Toda theories \cite{dunne1}, in the noncommutative scenario
the generalization of Toda theories proposed in \cite{kimeon} were
expressed as systems of first order equations which could not be
reduced to coupled second order equations in general.

By the other side the self-dual equations for Chern-Simons
solitons on nc space can be related to the equation of the $U(N)$
nc chiral model, which apparently can be also solved by a nc
extension of the uniton method of Uhlenbeck \cite{uhlenbeck} as
stated in \cite{kimeon}. In this way solutions to the proposed nc
Toda theories can be explicitly constructed \cite{kimeon}.
Particulary in \cite{kimeon} was constructed the simplest solution
to the nc generalization of Liouville model as proposed in that
work.

In this section we would like to make contact between our nc
extensions of the abelian and abelian affine Toda models
(\ref{todaeqnmotion}), (\ref{todaeqnaffinenc}) and the proposal
presented in \cite{kimeon} for these Toda models.
 We will see in the following that the
nc Leznov-Saveliev equations can be also obtained from the nc
Chern-Simons self-dual soliton equations. This gives the
possibility of obtaining nc extensions of Toda and affine Toda
models described by second order differential equations and at the
same time it can be possible to use the equivalence of the
self-dual equations to the nc chiral model equation to construct
solutions of the nc Toda models.

Consider the nc four dimensional (anti-) self-dual Yang-Mills
equations for a non-abelian gauge theory in euclidian space
\begin{eqnarray}
F_{12}=-F_{34}, \quad F_{13}=F_{24}, \quad F_{14}=-F_{23},
\label{sdym}
\end{eqnarray}
where $F_{ij}=\partial_i A_j-\partial_j A_i+[A_i,A_j]_{\star}$ is
the field strength. Considering the covariant derivatives defined
as $D_i=\partial_i+[A_i, \,\,]_{\star}$ and taking all fields to
be independent of $x_3$ and $x_4$, the equations (\ref{sdym})
reduce to
\begin{eqnarray}
F_{12}=-[A_3,A_4]_{\star}, \quad D_1 A_3=D_2 A_4, \quad D_1
A_4=-D_2 A_3, \label{reducedYM}
\end{eqnarray}
As the next step define the covariant derivatives as $D=D_1+iD_2$,
$\bar{D}=D_1-iD_2$ and the gauge fields as $A=A_1+iA_2$,
$\bar{A}=A_1-iA_2$ with the coordinates
$\tilde{z}=\frac{x_1-ix_2}{2}$,
$\bar{\tilde{z}}=\frac{x_1+ix_2}{2}$ and the corresponding partial
derivatives $\tilde{\partial}=\partial_1+i\partial_2$,
$\bar{\tilde{\partial}}=\partial_1-i\partial_2$. If now we
identify $\Psi=\sqrt{\kappa}(A_3-iA_4)$ (\ref{reducedYM})
transforms to the Chern-Simons self-dual equations,
 \begin{eqnarray}
\bar{D} \Psi=\bar{\tilde{\partial}}\Psi+[\bar{A},\Psi]_{\star}&=& 0, \nonumber \\
F_{+-}=\bar{\tilde{\partial}}A-\tilde{\partial}\bar{A}+[\bar{A},A]_{\star}&=&\frac{1}{k}[\Psi^{\dag},
\Psi]_{\star}, \label{selfdualchsimons}
\end{eqnarray}
where we have considered that $A_i^{\dagger}=A_i$ for $i=1\dots
4$. The solutions of this system yield static, minimum (zero)
energy configurations of a model in $2+1$ dimensions describing
charged scalar fields $\Psi$ with nonrelativistic dynamics,
minimally coupled to $U(N)$ gauge fields $A_{\mu}$ ($\mu=t,x,y$)
with Chern-Simons dynamics in the adjoint representation. The
scalar fields $\Psi$ and the gauge fields $A_{\mu}$ take values in
the same representation of the gauge Lie algebra. So, the nc
self-dual Chern-Simons equations in two dimensions can be obtained
through a dimensional reduction from the nc self-dual Yang-Mill
equations in four dimensions. In the following we will see how the
nc Leznov-Saveliev equations (\ref{Lez-Savnc}) can be also
obtained from (\ref{selfdualchsimons}). For this purpose let us
consider that the gauge fields are expressed as
\begin{eqnarray}
A &=&G^{-1}\star \tilde{\partial}  G,\label{A}\\
\bar{A}&=&-A^{\dag},
\end{eqnarray}
where $G$ is an element of the complexification of the gauge group
$G$. Suppose we can decompose $G$ as
 \be
G=H \star U,
 \ee
where $H$ is hermitian and $U$ is unitary. Let us now take our
original variables $z, \bar{z}$ which are related to the variables
$\tilde{z}, \bar{\tilde{z}}$ through $ z=2\tilde{z}$,
$\bar{z}=2\bar{\tilde{z}}$ and with $x_2 \rightarrow ix_2=-x$ and
$x_1=t$.
 The field strength is then expressed as
 \be
 F_{+-}=4U^{-1}\star H \star \bar{\partial}(H^{-2}\star\partial
H^2)\star H^{-1}\star U, \ee where $H^{2}=H \star H$ and
$H^{-2}=H^{-1}\star H^{-1}$.
 The solution of the
self-duality equation $\bar{D} \Psi=0$ is trivially: \be
\Psi=\sqrt{k} G \star \Psi_0(z)\star G^{-1}, \label{psi} \ee for
any $\Psi_0(z)$. Inserting this solution in the other self-duality
equation (\ref{selfdualchsimons}) yields the equation for $H$:
 \be
4\bar{\partial}(H^{-2}\star \partial H^2)=-\Psi_0 \star H^{-2}
\star \Psi_0^{\dag} \star H^2+H^{-2}\star \Psi_0^{\dag} \star H^2
\star \Psi_0. \ee Consider now that $H^2=B$, i. e. an element of
the zero grade subgroup and $\Psi_0=2\epsilon_-$, i.e. the
generator of grade $- 1$ which satisfy
$\epsilon_-^{\dag}=\epsilon_+$. Then the previous equation is
written like the nc Leznov-Saveliev equation (\ref{Lez-Savnc}).
This procedure is a nc extension of an alternative way for
deriving the Toda models from the Chern-Simons self-dual equations
presented in \cite{dunne1} and which differs from the ansatz used
in \cite{kimeon}. Nevertheless both approaches can be mapped one
into the another. For instance in \cite{kimeon} was proposed for
$U(N)$ the ansatz
\begin{eqnarray}
A=\mathrm{diag}(E_1,E_2,...,E_N), \quad
\tilde{\Psi}_{ij}=\delta_{i,j-1}h_i, \quad i=1...N-1, \label{kime}
\end{eqnarray}
which after introducing in (\ref{selfdualchsimons}) for $\Psi
\rightarrow \tilde{\Psi}$ leads to a system of coupled first order
equations for the $E_1, \dots, E_N$ fields and for the $h_1,\dots
,h_{N-1}$ fields (see ($4.2$) and ($4.3$) of \cite{kimeon}). From
(\ref{A}) we see that $A$ is expressed in terms of first order
derivatives
\begin{eqnarray}
A=2U^{-1}\star H^{-1}\star \partial H \star U+2U^{-1}\star
\partial U \label{AA},
\end{eqnarray}
and
 \be
 \Psi=2\sqrt{k}U^{-1}\star H \epsilon_{-}\star H^{-1} \star U. \label{psi1}
 \ee
With this choice the fields $h_1\dots h_{N-1}$ and $E_{1}\dots
E_{N}$ are no longer independent. For $U(N)$ we can also take the
constant generators as $\epsilon_{\pm}=\sum_{i=1}^{n-1}\mu_iE_{\pm
\alpha_i}$ and if we consider that $G=G_0$, i.e. the zero grade
subgroup it is possible to choose the unitary matrix $U$ in
(\ref{AA}) and (\ref{psi1}) as the identity matrix. In this way
$A$ in (\ref{AA}) will be also a diagonal matrix. More
specifically if $B$ is represented by the diagonal matrix \be
B=\mathrm{diag}(g_1, g_2,\dots g_N), \ee with
$g_i=e_{\star}^{\phi_i}$ for $i=1 \dots N$, then
\begin{eqnarray}
\Psi_{ij}=\delta_{i,j+1}g_{i+1}^{1/2}\star g^{-1/2}_{i},
\end{eqnarray}
where by $g_{i}^{1/2}$ is understood the function such that
$g_{i}^{1/2}\star g_{i}^{1/2}=g_i $. In this way we can relate
$\Psi=\tilde{\Psi}^{\dag}$, then
\begin{eqnarray}
\bar{h}_i&=&g_{i+1}^{1/2} \star g^{ -1/2}_{i}, \quad
\mathrm{for} \quad i=1\dots N-1, \nonumber \\
E_i&=&g^{-1/2}_{i}\star \partial g^{1/2}_{i}, \quad \mathrm{for}
\quad i=1\dots N. \label{relations1}
\end{eqnarray}
For the affine case the ansatz considered in \cite{kimeon} was
\begin{eqnarray}
A&=&\mathrm{diag}(E_1,E_2,...E_N),  \\
\Psi_{ij}&=&\delta_{i,j-1}h_i, \quad \mathrm{for} \quad i=1...N-1,
\quad \mathrm{except} \quad \mathrm{for} \quad \Psi_{N1}=h_N
\nonumber \label{kimeaffine}.
\end{eqnarray}
Here again we can established relations analogs to
(\ref{relations1}) using (\ref{AA}) and (\ref{psi1}), but now
remembering that $\epsilon_{\pm}=\sum_{i=1}^{N-1}\mu_iE_{\pm
\alpha_i}^{(0)}+m_0 E^{(\pm1)}_{\mp\psi}$. The relations obtained
are essentially (\ref{relations1}), except for the extra component
$\Psi_{N1}=h_N=g_{N} \star g_1^{-1}$ coming from the extra affine
generator. In all this process we have combined the constant
parameters in such a way that $2 \sqrt{k} \mu_i=1$ for $i=1 \dots
N-1$ and $2 \sqrt{k} m_0=1$.

We have seen how this way allows to define the nc abelian Toda
field theories as second order differential equations from the nc
self-dual Chern-Simons equations. Moreover using the relation of
the nc Chern-Simons to the nc principal chiral model \cite{kimeon}
will be possible to construct the solutions to the nc self-dual
equations (\ref{selfdualchsimons}) and in this way to the nc Toda
models from the solutions of the nc chiral model. In the ordinary
commutative case there is a well established procedure to
construct the solutions of the chiral model equation with have
finite energy called the Uniton method \cite{uhlenbeck}. In
\cite{kimeon} was conjectured the extension of this method to the
nc plane and was explicitly constructed an specific solution (the
simplest) to the nc Liouville model. In a forthcoming work
\cite{sdchsyo} we will present how solutions of our nc extension
of Toda models can be obtained by means of this procedure and the
relations (\ref{AA}) and (\ref{psi1}). By the other side the
possibility of having the nc extensions as second order
differential equations with their corresponding actions it is more
convenient from the quantization point of view.


\section{Conclusions}

At the end of \cite{us} was expressed our intention of extending
to the nc plane the affine Toda field theories, generalizing in
this way the nc sine-Gordon model. With this work we have
accomplished this purpose. More specifically we have shown how the
nc Leznov-Saveliev equations, proposed in \cite{us}, are obtained
as the equations of motion of a constrained $WZNW_{\star}$ model
as well as of a constrained two-loop $WZNW_{\star}$. Starting from
these equations, we have extended to the noncommutative plane not
only the $\widetilde{{\cal GL}}(n)$ abelian affine Toda theories
but also the abelian Conformal Toda theories associated to the
algebra ${\cal GL}(n)$. The actions from which the nc equations of
motion of the models can be derived were also presented. As
particular examples the nc Liouville and nc sinh/sine-Gordon have
been discussed. We have seen how the zero grade subgroup in the nc
scenario looses the abelian cha\-rac\-ter. Due to that one can
choose alternative parameterization schemes that will lead to
equivalent nc extensions of the same model. These two-dimensional
theories (nc Liouville and nc sinh/sine-Gordon) can be also
obtained from four-dimensional nc SDYM in the Yang formulation
through a suitable dimensional reduction, as we have shown.
Furthermore, we have also seen that in general the nc
Leznov-Saveliev equations can be obtained from the nc (anti-)SDYM
by a dimensional reduction via the nc self-dual Chern-Simons
equations. This gives another example in favor of the validity of
the Ward conjecture on nc space-time \cite{ward}.

 The construction scheme that we have proposed gives the
 possibility of extending the
mo\-dels directly in two dimensions without apparently loosing the
integrability properties of the original field theory. The crucial
point is that the deformation must be done in a consistent way,
respecting the {\it algebra-group} relation, what means that we
must extend the group and its corresponding algebra.

We have explicitly studied the relation of our proposal to
previous ones es\-ta\-bli\-shing in this way the connection with
seemingly up to now disconnected versions. Our approach where the
nc Leznov-Saveliev equations play a crucial role allows to
establish these relations in a relatively simple way and
additionally it gives a general framework where many of the
previous proposals are included as particular cases. Besides that,
our scheme permits the formulation of these theories in form of
action principles what is crucial when quantization is intended.

 Of course, there are still several interesting directions to pursue in future research. Among
 them to investigate the full integrability of the nc theories
presented which means to construct
 the conserved charges and to have a more thorough understanding of their influence
 on the properties of the S-matrix. The symmetries and
the multisolitons solutions of the affine
 models proposed still require a deeper study. Most of the difficulties found in the investigation
 of these topics are related to the fact that when time is a noncommuting coordinate
 there is not a nc analog of the Noether theorem. By the other side, since the actions in the nc setup
 include infinite
 time derivatives is controversial the definition of the conjugate momenta
 and in the same way of the corresponding Hamiltonian, Poisson
 brackets and r-matrices, for example. We hope that the
 current investigation on this new area of integrability on nc
 spaces will shed more light on these subjects in the near future.

\section*{Acknowledgments}

 I thank the Physics Department-UFMT
 for the kind hospitality. I am very grateful to H. Blas for useful
discussions and to M. Moriconi for the ideas shared at the very
initial stage of this work. This work has been
 supported by CNPq-FAPEMAT.

\section{Appendix A}

Let us introduce here some of the algebraic structures used in
this paper. The Lie algebra ${\cal SL}(n)$ in the Chevalley basis
is defined through the commutation relations:
\begin{flushleft}
\hspace{2.5cm} $[h_i,h_j]= \quad 0$, \\
\hspace{2.2cm} $[h_i,E_{\alpha_j}] = \sum_{b=1}^{n-1}
m_b^{\alpha_j}k_{bi} E_{\alpha_j}$, \\
\end{flushleft}
\begin{equation}
 [E_{\alpha_i}, E_{\alpha_j}]= \left\{ \begin{array}{ccc}
                             \sum_{b=1}^{n-1} l_b^{\alpha_i}h_{b},
                              & \quad \mathrm{if} \quad
\alpha_i+\alpha_j &=0, \quad \quad \quad \quad\\

\varepsilon(\alpha_i,\alpha_j)E_{\alpha_i+\alpha_j},
                           &
                           \quad
                            \mathrm{if} \quad \alpha_i+\alpha_j &
                            \mathrm{is} \quad \mathrm{a} \quad
                            \mathrm{root}, \\
                            0 & \quad  \mathrm{otherwise}, &
                                    \end{array} \right.
\end{equation}
where $\varepsilon(\alpha_i,\alpha_j)$ are constant such that
$\varepsilon(\alpha_i,\alpha_j)=-\varepsilon(\alpha_j,\alpha_i)$,
$k_{ij}$ is the Cartan matrix $k_{ij}=\frac{\alpha_i \cdot
\alpha_j}{\alpha_j^2}, \quad i=1 \dots n-1$, with $\alpha_i,
\alpha_j$ in this case simple roots. For any root we have that
$\frac{\alpha_i}{\alpha_i^2}=\sum_{b=1}^{n-1}
l_b^{\alpha_i}\frac{\alpha_b}{\alpha_b^2} \quad \mathrm{and} \quad
\alpha_i=\sum_{b=1}^{n-1}m_b^{\alpha_i}\alpha_b$. The bilinear
form
\begin{eqnarray}
Tr(h_i h_j)&=&k_{ij}, \nonumber \\
Tr(E_{\alpha_i}E_{\alpha_j})&=&\frac{2}{|\alpha_i|^2}\delta_{\alpha_i+\alpha_j,0}
\quad \mathrm{for} \, \, \mathrm{any} \,\, \mathrm{root}, \nonumber \\
Tr(E_{\alpha_i} h_j)&=&0,
\end{eqnarray}
is also introduced. The loop algebra $\widetilde{{\cal SL}}(n)$ is
the Lie algebra of traceless matrices with entries which are
Laurent polynomials in $\lambda$
\begin{equation}
\widetilde{{\cal SL}}(n)=C(\lambda,\lambda^{-1})\otimes {\cal
SL}(n).
\end{equation}
The structure of the Lie algebra is introduced by the relation
\begin{equation}
[\lambda^n\otimes T_i, \lambda^m \otimes T_j]=\lambda^{n+m}\otimes
f_{ijk}T_k,
\end{equation}
where $m,n \in \mathbb{Z}$ and the elements of the form $1 \otimes
T_i$ are identified with the algebra ${\cal SL}(n)$ which is a
subalgebra of $\widetilde{{\cal SL}}(n)$. In this sense we can
write $\lambda^n \otimes T_i$ as $\lambda^n T_i$.

The derivation $d=\lambda \frac{d}{d\lambda}$ generator included
in the grading operators acts as  \be
[d,E_{\alpha_i}^{(n)}]=nE_{\alpha_i}^{(n)}, \quad [d,h_i^{(n)}]=n
h_i^{(n)}. \ee We have also considered the symmetric bilinear form
 \be Tr
(h_i^{(m)}h_j^{(n)})=k_{ij}\delta_{m+n,0},\ee \be Tr
(E_{\alpha_i}^{(m)}E_{\alpha_j}^{(n)})=\frac{2}{|\alpha_i^2|}\delta_{\alpha_i+\alpha_j,0}\delta_{m+n,0},
\ee and that $|\alpha_i|^2=2$ for simple roots.


\begin{thebibliography}{99}

\bibitem{stringnc}
Alain Connes, Michael R. Douglas and Albert Schwarz, JHEP {\bf
9802} (1998)
003; \\
Michael R. Douglas and Chris Hull, JHEP {\bf 9802} (1998)
008; \\
Nathan Seiberg and Edward Witten, JHEP {\bf 9909} (1999) 032.
\bibitem{dimakis}
Aristophanes Dimakis and Folkert Muller-Hoissen , Int. J. Mod.
Phys. {\bf B14} (2000) 2455-2460.
\bibitem{integrablenc}
Aristophanes Dimakis and Folkert Muller-Hoissen,
[arXiv:hep-th/0007015], [arXiv: hep-th/0007074], Lett.Math.Phys.
{\bf 54} (2000) 123-135, J.Phys. {\bf A37} (2004) 4069-4084,
J.Phys. {\bf A37} (2004) 10899-10930;
\\
Stefano Profumo, JHEP {\bf 0210} (2002) 035; \\
Masashi Hamanaka, J.Math.Phys.{\bf 46} (2005) 052701.
\bibitem{integrablenc2}
M. Legare, [arXiv:hep-th/0012077], J. Phys. {\bf A35} (2002) 5489; \\
Masashi Hamanaka and Kouichi Toda, J.Phys. {\bf A36} (2003)
11981-11998, Phys.Lett. {\bf A316} (2003) 77-83.
\bibitem{GP1}
 M. T. Grisaru and S. Penati, Nucl. Phys. {\bf B565} (2003) 250-276.
\bibitem{us}
I. Cabrera-Carnero and M. Moriconi, Nucl.Phys. {\bf B673} (2003)
437-454; see also [arXiv:hep-th/0303168], PRHEP-unesp2002/{\bf
028}.
\bibitem{GP2}
Marcus T. Grisaru, Liuba Mazzanti, Silvia Penati and Laura
Tamassia, JHEP {\bf 057} (2004) 0404.
\bibitem{lechtenfeld}
Olaf Lechtenfeld, Liuba Mazzanti, Silvia Penati, Alexander D.
Popov and Laura Tamassia, Nucl.Phys. {\bf B705} (2005) 477-503.
\bibitem{kimeon}
Ki-Myeong Lee, JHEP {\bf 0408} (2004) 054.
\bibitem{zuevsky}
A. Zuevsky, J. Phys. A: Math. Gen. {\bf 37} (2004) 537-547.

\bibitem{GM}
Jaume Gomis and Thomas Mehen, Nucl. Phys. {\bf B591} (2000)
265-276.
\bibitem{BDFP}
D.~Bahns, S.~Doplicher, K.~Fredenhagen and G.~Piacitelli, Phys.
Lett. {\bf B533} (2002) 178-181.
\bibitem{CLZ}
Chong-Sun Chu, Jerzy Lukierski and Wojtek J. Zakrzewski,  Nucl.
Phys. {\bf B632} (2002) 219-239.
\bibitem{D}
Patrick Dorey, [arXiv:hep-th/9810026].
\bibitem{witten}
E. Witten, Commun. Math. Phys. {\bf 92} (1984) 455-472; \\
J. Wess and B. Zumino, Phys. Lett. {\bf B37} (1971) 95; \\
S. P. Novikov, Sov. Math. Dock. {\bf 24} (1981) 222.
\bibitem{balog}
J. Balog, L. Feher, L. O'Raifeartaigh, P. Forgas and A. Wipf,
Phys. Lett. {\bf B227} (1989) 214, Phys. Lett. {\bf B244} (1990)
435-441, Ann. Phys. {\bf 203} (1990) 76-136.

\bibitem{aratyn}
 H. Aratyn, L.A. Ferreira, J. F. Gomes and H.
Zimmerman, Phys. Lett. {\bf B254} (1991) 372-380.
\bibitem{LezSav}
A. N. Leznov and M. V. Saveliev, Commun. Math. Phys. {\bf 89}
(1983) 59.
\bibitem{olive}
D. I. Olive and N. Turok, Nucl. Phys. {\bf B215} (1983) 470.

\bibitem{bonora}
L. Bonora, M. Martellini and Y. Z. Zhang, Phys. Lett. {\bf B253}
(1991) 373-379.
\bibitem{babelon}
O. Babelon and L. Bonora, Phys. Lett {\bf B244} (1990) 220-226.

\bibitem{pinzul}
A. Pinzul and A. Stern, [arXiv:hep-th/0406068].

\bibitem{constantinidis}
C. P. Constantinidis, L. A. Ferreira, J. F. Gomes and A. H.
Zimerman, Phys. Lett. {\bf B298} (1993) 88-94.

\bibitem{popov3}
Olaf Lechtenfeld, Alexander D. Popov and B. Spending, Phys. Lett
{\bf B507} (2001) 317-326.

\bibitem{Toda}
M. Toda, J. Phys. Soc. Jap. {\bf 22} (1967) 431.

\bibitem{olive1}
D. I. Olive, N. Turok and J.W.R.Underwood, Nucl. Phys. {\bf B409}
(1993) 509-546.

\bibitem{wznwnc}
E. F. Moreno and F. A. Shaposnik, JHEP {\bf 0003} (2000) 032.

\bibitem{reviews}
Michael R. Douglas and Nikita A. Nekrasov, Rev. Mod. Phys. {\bf
73}
(2001) 977-1029; \\
Richard J. Szabo, Phys.Rept. {\bf 378} (2003) 207-299.
\bibitem{dunne1}
Gerald V. Dunne, R. Jackiw, So-Young Pi and Carlo A. Trugenberg,
Phys. Rev {\bf D43} (1991) 1332-1345.
\bibitem{dunne2}
G. V. Dunne, Commun. Math. Phys. {\bf 150} (1992) 519-535.
\bibitem{dunne3}
G. V. Dunne, [arXiv:hep-th/9410065].
\bibitem{uhlenbeck}
K. Uhlenbeck, J. Dif. Geom. {\bf 30} (1989) 1.
\bibitem{save-ver}
M. V. Saveliev and A. M. Vershik, Commun. Math. Phys. {\bf 126}
(1989) 367.

\bibitem{bilal}
Adel Bilal and Jean-Loup Gervais, Phys. Lett. {\bf B206} (1988)
412, Nucl. Phys. {\bf B314} (1989) 646, Nucl. Phys. {\bf B318}
(1989)
579; \\
O. Babelon, Phys. Lett. {\bf B215} (1988) 523.

\bibitem{moyal}
J. E. Moyal, Proc. Cambridge Phil. Soc. {\bf 45} (1949) 99.
\bibitem{parvizi}
Amir Masoud Ghezelbash and Shahrokh Parvizi, Nucl.Phys. {\bf B592}
(2001) 408-416.

\bibitem{sakakibara}
M. Sakakibara, J. Phys. A: Math. Gen. {\bf 37} (2004) 599-604.
\bibitem{ueno}
K. Ueno and K. Takasaki, Proc. Japan Acad. Ser {\bf A59} (1983)
167-170; Proc. Japan Acad. Ser {\bf A59} (1983) 215-218.

\bibitem{emilio}
 E.P. Gueuvoghlanian, [arXiv:hep-th/0105015].

\bibitem{hollowood}
T. J. Hollowood,  Nucl.Phys. {\bf 384B} (1992) 523-540.

\bibitem{dressing}
V. E. Zakharov and V. A. Mikhailov, Sov. Phys. JETP {\bf 47}
(1978) 1017-1027; \\
V. E. Zakharov and A. B. Shabat, Funct. Anal. Appl. {\bf 13}
(1979) 166; \\
P. Forgacs, Z. Horvath and L. Palla, Nucl.Phys. {\bf B229} (1983)
77 .
\bibitem{harold}
H. Blas, H. L. Carrion and M. Rojas, JHEP {\bf 0503} (2005) 037.
\bibitem{ward}
R. S. Ward, Phil. Trans. R. Soc. Lond. {\bf A315} (1985) 451.
\bibitem{popov2}
Olaf Lechtenfeld and Alexander D. Popov, JHEP {\bf 0111} (2001)
040; Phys. Lett. {\bf B523} (2001) 178-184.
\bibitem{takasaki}
Kanehisa Takasaki, J.Geom.Phys. {\bf 37} (2001) 291-306.
\bibitem{mo}
Mo-Lin Ge, Lai Wang and Yong-Shi Wu, Phys. Lett {\bf B335} (1994)
136-142.
\bibitem{sdchsyo}
I. Cabrera-Carnero, in preparation.




\end{thebibliography}
\end{document}